\documentclass[final,5pt,twocolumn]{elsarticle}
\usepackage{graphicx}
\usepackage{amsmath,amssymb}
\usepackage[amssymb]{SIunits}
\usepackage{SIunits}
\usepackage{xspace}
\usepackage{color}
\usepackage[colorlinks,urlcolor=blue]{hyperref}
\usepackage[all]{hypcap}
\biboptions{sort&compress}
\newcommand*{\doi}[1]{\href{http://dx.doi.org/\detokenize{#1}}{\detokenize{#1}}}

\def\vec#1{\ensuremath{{\bf #1}}\xspace}

\def\Poincare{Poincar\'e\xspace}

\def\bfield#1{\ensuremath{\vec{B}_{\textnormal{#1}}}\xspace}
\def\dB{\ensuremath{\delta\vec{B}}\xspace}
\def\psiN{\ensuremath{\psi}\xspace}
\def\psisepx{\ensuremath{\psi_{\textnormal{sepx}}}\xspace}
\def\psiaxis{\ensuremath{\psi_{\textnormal{axis}}}\xspace}

\def\nsym{\ensuremath{n_{\textnormal{sym}}}\xspace}

\begin{document}

\begin{frontmatter}
\title{FLARE: field line analysis and reconstruction for 3D plasma boundary modeling}

\author[1]{H. Frerichs\corref{author}}
\cortext[author] {Corresponding author.\\\textit{E-mail address:} h.frerichs@wisc.edu}

\address[1]{University of Wisconsin - Madison, Department of Engineering Physics, Madison, WI 53706, USA}

\journal{arXiv}

\begin{abstract}
The FLARE code is a magnetic mesh generator that is integrated within a suite of tools for the analysis of the magnetic geometry in toroidal fusion devices.
A magnetic mesh is constructed from field line segments and permits fast reconstruction of field lines in 3D plasma boundary codes such as EMC3-EIRENE.
Both intrinsically non-axisymmetric configurations (stellarators) and those with symmetry breaking perturbations of an axisymmetric equilibrium (tokamaks) are supported.
The code itself is written in Modern Fortran with MPI support for parallel computing, and it incorporates object-oriented programming for the definition of the magnetic field and the material surface geometry.
Extended derived types for a number of different magnetohydrodynamic (MHD) equilibrium and plasma response models are implemented.
The core element of FLARE is a field line tracer with adaptive step-size control, and this is integrated into tools for the construction of \Poincare maps and invariant manifolds of X-points.
A collection of high-level procedures that generate output files for visualization is build on top of that.
The analysis modules are build with Python frontends that facilitate customization of tasks and/or scripting of parameter scans.
\end{abstract}

\end{frontmatter}

\section{Introduction} \label{sec:introduction}

The plasma boundary in magnetic confinement devices must accommodate a hot plasma on the one side with material surfaces on the other.
This includes a number of challenges from anisotropic electron heat conductivity (plasma physics) to interactions between charged and neutral particles (atomic and molecular physics).
The former implies that exhaust from the confined plasma travels along magnetic field lines to a rather small area on so called divertor targets, and the latter comes from recombination of electrons and ions on material surfaces (or in the divertor volume if it is cold enough) and the re-ionization of the resulting neutral particles when they are released back into the plasma.
Detachment of the plasma from material surfaces through strong plasma-neutral interaction, impurity radiation and volumetric recombination is the main operating scenario of the ITER divertor (ITER physics basis \cite{ITERPhysicsExpertGroupOnDivertor1999}).
Both reduced (two-point \cite{Mandrekas1996, Sugihara1997, Stangeby2000, Kotov2009, Siccinio2016, Stangeby2018} or 1D \cite{Hutchinson1994, Nakazawa2000, Goswami2001, Havlickova2011, Lipschultz2016}) and high fidelity (2D \cite{Rognlien1992, Reiter1992, Simonini1994, Smith1995, Schneider2006, Kawashima2006, Wiesen2015} or 3D \cite{Runov2001, Feng2004, Feng2017, Frerichs2021}) models are widely-used for analysis of the plasma boundary, and they all have in common that the plasma is treated as a fluid-in-a-magnetic-field problem rather than a full magnetohydrodynamics (MHD) one.
In the following we will focus on the {\em magnetic geometry} (i.e. properties of the magnetic field in combination with material surfaces), either as input for plasma boundary models or for interpretation of experimental measurements.

The FLARE code is a magnetic mesh generator that is integrated within a suite of tools for the analysis of the magnetic geometry.
It is intended for configurations without axisymmetry (continuous symmetry with respect to the toroidal angle), such as intrinsically non-axisymmetric stellarators and tokamaks with symmetry breaking resonant magnetic perturbations (RMPs) \cite{Evans2004, Evans2015} for control of edge localized modes (ELMs).
It has been developed in support of 3D plasma boundary modeling, and has already found applications in DIII-D \cite{Frerichs2014b, Frerichs2015b}, NSTX-U \cite{Poster-APS2015-Frerichs, Frerichs2016}, W7-X \cite{ISHW2017-Effenberg, Lore2018, Effenberg2019a, Boeyaert2023}, MAST \cite{Waters2018}, ITER \cite{Frerichs2019, Frerichs2020, Frerichs2021a}, Heliotron J \cite{Matoike2019, Matoike2021}, Wistell-A \cite{HSFE2019-Frerichs, Bader2020}, KSTAR \cite{Frerichs2023}, CTH \cite{Garcia2023} and QFCS \cite{Poster-PET2023-BLiu} - but often without reference.
The purpose of this manuscript is to provide long overdue documentation of FLARE itself and its capabilites.

\begin{figure}
\centering
\includegraphics[width=80mm]{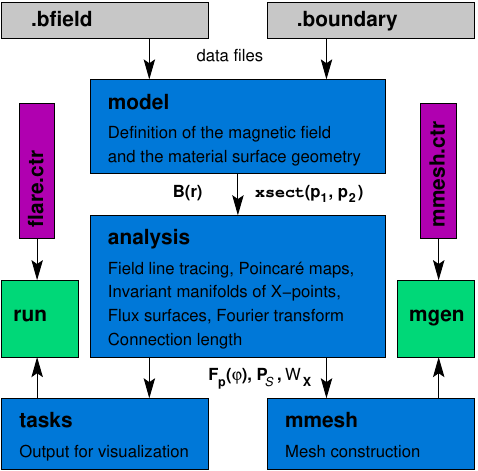}
\caption{Structure of the FLARE code: library modules written in Fortran with Python frontends (blue), configuration files for the numerical model (gray), command line interfaces (green) and their control files (magenta).}
\label{fig:code_structure}
\end{figure}

The code itself is written in Modern Fortran with MPI support for parallel computing.
The structure of the code is highlighted in figure \ref{fig:code_structure}.
Python frontends are provided for customization of analysis tasks and/or scripting of parameter scans.
Object-oriented programming is used for the definition of the magnetic field and the material surface geometry.
Extended derived types (classes in Python) for a number of different magnetohydrodynamic (MHD) equilibrium and plasma response models are implemented.
As such, FLARE is not a physics model itself, but requires data files (from other codes) as input.
The definition of the {\em numerical model} which provides functions for the magnetic field $\vec{B}(\vec{r})$ and for boundary intersection checks $\texttt{xsect}(\vec{p}_1, \vec{p}_2)$ is summarized in section \ref{sec:model}.
Cylindrical coordinates $(r,z,\varphi)$ are used in FLARE, and the toroidal angle $\varphi$ increases in counter-clockwise direction as seen from above.

The low-level analysis tools are implemented as derived types / classes, and can be integrated into more advanced tasks.
They are independent of the selected implementation of the magnetic field and boundary geometry.
Central to FLARE is a field line tracer with adaptive step-size control, and this is integrated into tools for the construction of \Poincare maps and invariant manifolds of X-points.
Additional tools support the construction of flux surfaces and Fourier transform of perturbation fields.
The analysis module and a collection of high-level tasks are described in section \ref{sec:analysis}.
The latter store results in output files for visualization, and they can also be executed through a command line interface with control files.
More details can be found in the online FLARE user manual \cite{FLARE}.

In plasma boundary modeling, Monte Carlo methods require rapid evaluation of field line segments.
Significant speedup over numerical integration can be achieved based on interpolation within a finite magnetic flux-tube mesh (in the following referred to as magnetic mesh) in combination with a reversible mapping between two adjacent flux-tubes \cite{Feng2005}.
A magnetic mesh generator for the 3D plasma boundary code EMC3-EIRENE \cite{Feng2004, Feng2017} is described in section \ref{sec:mmesh}.

\section{Model definition} \label{sec:model}

The foundation of all analysis tools is the numerical implementation of the magnetic field and the material surface geometry.
The purpose of this {\it numerical model} is to provide functions for the magnetic field $\vec{B}(\vec{r})$ and for intersection checks $\texttt{xsect}(\vec{p}_1, \vec{p}_2)$ between field line segments $\vec{p}_1 \rightarrow \vec{p}_2$ and material surfaces (divertor targets, limiters, ...). 
Those functions can be used by the analysis tools regardless of the selected implementation.
The numerical model is selected at run-time through configuration files \texttt{.bfield} and \texttt{.boundary} (local or from a database), or customized as part of a Python script.
Available implementations of the magnetic field are described below in section \ref{sec:bfield}, followed by implementations of the boundary geometry in section \ref{sec:boundary}.

\subsection{Magnetic field} \label{sec:bfield}

The magnetic field can be defined as a superposition of an equilibrium field and a number of (optional) perturbation fields:

\begin{equation}
\vec{B}(\vec{r}) \, = \, \bfield{equi}(\vec{r}) \, + \, \sum_{i=1}^{m} \bfield{3d,i}(\vec{r}). \label{eq:Bequi+perturbation}
\end{equation}

The distinction between \bfield{equi} and the sum of \bfield{3d,i} is irrelevant for field line tracing, but it supports additional analysis dedicated to perturbation fields in tokamaks (e.g. Fourier transform).
The equilibrium field is supplemented by a definition of the magnetic axis $\vec{r}_0(\varphi)$, which can be non-axisymmetric for stellarators.
Coordinate transformations for the circular poloidal angle $\theta(\vec{r})$ and minor radius $\rho(\vec{r})$ are provided.
Specifics of axisymmetric equilibrium field for tokamaks, perturbation fields, and non-axisymmetric equilibrium fields for stellarators are described below.

\subsubsection{Axisymmetric equilibrium fields} \label{sec:equi2d}

In an axisymmetric ideal MHD equilibrium, the poloidal flux function $\Psi(r,z)$, plasma pressure $p(\Psi)$ and toroidal field (and net poloidal current) function $g(\Psi)$ are related by the Grad-Shafranov equation \cite{Grad1958, Shafranov1966}, a two-dimensional, non-linear, elliptic partial differential equation.
The magnetic field is given by

\begin{equation}
\bfield{equi2d}(r,z) \, = \, \frac{1}{r} \, \nabla \Psi \, \times \, \vec{e}_\varphi \, + \, \frac{g(\Psi)}{r} \, \vec{e}_\varphi. \label{eq:equi2d}
\end{equation}

Implementations of (\ref{eq:equi2d}) are extended from an abstract type \texttt{equi2d} which provides common parameters and functions.
The former include the location of X-points, which can be detected automatically from a grid search for critical points within the equilibrium domain.
The latter include evaluation of the normalized poloidal flux

\begin{equation}
\psiN(r,z) \, = \, \frac{\Psi \, - \, \Psi_{\textnormal{axis}}}{\Psi_{\textnormal{sepx}} \, - \, \Psi_{\textnormal{axis}}} \label{eq:psiN}
\end{equation}

with respect to the flux on the magnetic axis $\Psi_{\textnormal{axis}}$ and on the separatrix $\Psi_{\textnormal{sepx}}$, and an inverse coordinate transformation $\texttt{rzcoords}(\psiN, \theta)$ which computes the corresponding $(r,z)$ from a 1D root finder along the direction $\theta$ from the magnetic axis $\vec{r}_0$.

An analytic solution \cite{Cerfon2010} for simple pressure and poloidal current profiles with arbitrary aspect ratio, elongation, triangularity and optional X-point location is implemented for academic purposes.
For practical applications, however, a 2D B-spline interpolation $\Psi_{\textnormal{bspline2d}}(r,z)$ is implemented for data on a rectangular grid, either from forward modeling (CHEASE \cite{Lutjens1996}, CORSICA \cite{Crotinger1997}, DIVA, FIESTA) or from equilibrium reconstruction (EFIT \cite{Lao1985, Lao1990}, LRDFIT).
Depending on the data model, the toroidal field function is either implemented as an interpolating cubic spline $g_{\textnormal{interp}}(\psiN)$ in the confined region with constant extrapolation outside, or as a global constant value $g_0$.

\subsubsection{Perturbation fields} \label{sec:perturbations}

In tokamaks, symmetry breaking perturbation fields are intentionally imposed for control of ELMs by dedicated coils outside the plasma.
These perturbations are aligned with resonances at the plasma edge, and can cause a reduction of the pressure gradient in the H-mode pedestal region which improves stability.
For many years, the so called vacuum RMP approximation has been applied where the perturbation field inside the plasma is computed only from the coils outside the plasma.
This functionality is retained for reference (see \texttt{coilset} below).
However, plasma response effects (screening and amplification of components) are important in most present day tokamaks, and support for a number of different models is implemented in FLARE.
Linear combination of perturbation fields from different coil groups or toroidal mode numbers $n$ is supported (for external fields or linear MHD plasma response data).
Below is a summary of available perturbation field types:

\begin{itemize}
\item \texttt{coilset}: This is for fields from external coils.
Coils are approximated by a sequence of straight line segments that represent current filaments.
A compact Biot-Savart expression for the fields is implemented which is singular only on the segment itself \cite{Hanson2002}.
This offers a high degree of flexibility regarding the shape of coils, but it can become computationally expensive for configuration with very complex coils or for thick coils that require multiple loops of current filaments.

\item \texttt{bspline3d}: A 3D tensor product B-spline interpolation is implemented for a) the vector potential and b) the magnetic field on a regular cylindrical grid.
The advantage of the former is that it automatically satisfies $\nabla \, \cdot \, \vec{B} \, = \, 0$.

\item \texttt{interp}: This is a Cubic Hermite interpolation method for the magnetic field on a regular cylindrical grid.
In particular, this method enforces $\nabla \, \cdot \, \vec{B} \, = \, 0$ and $\nabla \, \times \, \vec{B} \, = \, 0$.
It is suitable only for fields from external coils, e.g. as a faster alternative for \texttt{coilset} (from which data files for interpolation can be generated).

\item \texttt{gpec}: The IPEC code \cite{Park2007a, Park2009, Park2010} computes free-boundary ideal perturbed equilibria for given pressure $p(\psi)$ and safety factor $q(\psi)$ profiles by solving the perturbed force balance equation.
The GPEC code \cite{Park2017} is a generalization which calculates the kinetic force balance with self-consistent neoclassical toroidal viscosity (NTV) torque.
Boundary conditions are applied on a control surface near the plasma boundary based on a virtual casing principle for external fields.

\item \texttt{marsf}: The MARS-F \cite{Liu2000, Liu2010} plasma response calculations based on a linearized resistive single-fluid MHD model in a flux coordinate system in toroidal geometry.
A Fourier expansion is used in the poloidal direction, and different toroidal mode numbers can be considered by independent calculations.
For divertor configurations, the coordinate system is extended into the scrape-off layer by smoothing the equilibrium near the X-point.

\item \texttt{m3dc1}: The M3D-C${}^1$ \cite{Jardin2007, Jardin2008, Ferraro2012} plasma response calculations are based on a resistive two-fluid MHD model in diverted, toroidal geometry.
Quintic triangular finite elements are used with constrained quintic terms in order to enforce $C^1$ continuity across element boundaries \cite{Jardin2004}.

\item \texttt{jorek}: The JOREK non-linear extended MHD code is based on bicubic B\'ezier finite elements and a toroidal Fourier expansion \cite{Czarny2008, Hoelzl2021}.
Elements are constrained for first order continuity in real space, except at the magnetic axis and in the vicinity of X-points.
\end{itemize}

\subsubsection{Non-axisymmetric equilibrium fields} \label{sec:equi3d}

Unlike tokamaks, stellarator configurations are non-axisymmetric by design as they rely primarily on complex external coils for magnetic confinement of the plasma.
A non-axisymmetric equilibrium is implemented in FLARE as

\begin{equation}
\bfield{equi3d}(\vec{r}) \, = \, \sum_{i=1}^{m} \bfield{3d,i}(\vec{r}) \label{eq:Bequi3d}
\end{equation}

which permits individual adjustment of coil groups similar to the implementation of perturbation fields in tokamaks.
In particular, 3D equilibrium specific implementations of the \texttt{coilset} and \texttt{interp} field types are available. 
The magnetic axis $\vec{r}_0(\varphi)$ for a non-axisymmetric equilibrium is implemented as a periodic cubic spline and can be constructed automatically by iterative approximation.
The following additional implementations are available for 3D equilibrium fields:

\begin{itemize}
\item \texttt{bmw}: This is an interface for option a) of the \texttt{bspline3d} field type for data files generated with BMW.
The BMW code (by M. Cianciosa) computes the magnetic vector potential on a regular cylindrical grid from a VMEC equilibrium by volume integral over the plasma.

\item \texttt{mgrid}: This is an interface for option b) of the \texttt{bspline3d} field type for data files generated with MAKEGRID \cite{MAKEGRID} or EXTENDER \cite{Drevlak2005}.
The latter calculates the extended magnetic field for VMEC and PIES equilibria through a virtual casing principle.

\item \texttt{hint}: The HINT code \cite{Suzuki2006, Suzuki2017} is a 3D MHD equilibrium solver which alternates between relaxation of the plasma pressure $p$ (at fixed $\vec{B}$) and relaxation of the magnetic field (at fixed $p$).
It uses cylindrical coordinates which permits to extend the simulation domain beyond the last closed flux surface.
\end{itemize}

\subsection{Boundary geometry} \label{sec:boundary}

The second part of the numerical model is the implementation of the boundary geometry.
This is for the purpose of intercepting magnetic field lines and does not require to keep track of material properties.
Two types of boundaries are supported: axisymmetric surfaces (\texttt{axisurf}) and non-axisymmetric ones with toroidal layout (\texttt{torosurf}).
Both types are described below.

The boundary can be defined as a combination of surface patches, and surface patches of either type can be combined.
Specifically, the intention here is to define a function

\begin{equation}
\texttt{xsect}(\vec{p}_1, \vec{p}_2) \rightarrow (t, k, \vec{u}) \label{eq:xsect}
\end{equation}

which checks if a field line segment from $\vec{p}_1$ to $\vec{p}_2$ is intercepted on a material surface.
If the intersection check (\ref{eq:xsect}) succeeds, then $t \, \in \, [0,1]$ defines the intersection point

\begin{equation}
\vec{p}_x \, = \, \vec{p}_1 \, + \, t \, (\vec{p}_2 - \vec{p}_1). \label{eq:px_linear}
\end{equation}

along the linearized field line segment
Also, the tuple $\vec{u}$ of 2D surface coordinates associated with the intersection point on the $k$-th surface patch is returned in (\ref{eq:xsect}).
This can be useful for visualization or further analysis of the strike point.

\subsubsection{Axisymmetric surfaces} \label{sec:axisurf}

A toroidally symmetric surface is created from a 2D curve in the R-Z plane that is rotated around the longitudinal axis of the cylindrical coordinate system.
This is known as surface of revolution.
The \texttt{axisurf} type is a polygonal representation

\begin{equation}
\vec{c}_j \, = \, (r_j, z_j), \qquad j = 0 \ldots M \label{eq:axisurf}
\end{equation}

with $M$ segments.
For the intersection point with a linearized field line segment, the $r$ and $z$ components of (\ref{eq:px_linear}) are combined with

\begin{equation}
(r_x, z_x) \, = \, \vec{c}_j \, + \, s \, (\vec{c}_{j+1} - \vec{c}_j), \qquad s \in [0,1] \label{eq:xsect2d}
\end{equation}

for the $j$-th polygon segment to form a system of two linear equations for $(s, t)$ which is straightforward to solve.
The third component of (\ref{eq:px_linear}) then determines $\varphi_x$.
The tuple of surface coordinates associated with the intersection point is defined as

\begin{equation}
\vec{u} = (\varphi_x \mod 2 \pi, \, j+s).
\end{equation}

\subsubsection{Toroidal quadrilateral elements}

For non-axisymmetric boundaries in toroidal configurations, it is useful to think of the rotated polygon segments as a special case of a doubly ruled surface in cylindrical coordinates.
The \texttt{torosurf} type accounts for the more general case of $M \times N$ doubly ruled quadrilateral surface patches with nodes

\begin{equation}
\label{eq:torosurf}
\begin{split}
\vec{c}_{ij} & = \, (r_{ij}, z_{ij}), \\
\varphi_{ij} & = \, \varphi_i, \qquad i = 0 \ldots N, \quad j = 0 \ldots M.
\end{split}
\end{equation}

Essentially, this is a sequence of $N+1$ polygons at different toroidal locations $\varphi_i$.
The intersection point with a field line segment can be expressed in surface coordinates as

\begin{subequations}
\begin{align}
(r_x, z_x) & = \, \vec{c}_{ij} \, + \, s_1 \, \vec{S}^b_{ij} \, + \, s_2 \, \vec{S}^c_{ij} \, + \, s_1 \, s_2 \, \vec{S}^d_{ij} \label{eq:xsect3d} \\
\varphi_x  & = \, \varphi_i \, + \, s_1 \, \left(\varphi_{i+1} \, - \, \varphi_i\right) \label{eq:xsect3d_phi}
\end{align}
\end{subequations}

with $s_1, s_2 \, \in \, [0,1]$.
Unlike (\ref{eq:xsect2d}), this includes an additional equation for $\varphi_x$.
The shape coefficients in (\ref{eq:xsect3d}) are determined by the 4 nodes of a quadrilateral element:

\begin{eqnarray}
\vec{S}^b_{ij} & = & \vec{c}_{(i+1) j} \, - \, \vec{c}_{ij} \notag \\
\vec{S}^c_{ij} & = & \vec{c}_{i (j+1)} \, - \, \vec{c}_{ij} \\
\vec{S}^d_{ij} & = & \vec{c}_{(i+1) (j+1)} \, - \, \vec{c}_{(i+1) j} \, - \, \vec{c}_{i (j+1)} \, + \, \vec{c}_{ij} \notag.
\end{eqnarray}

The surface coordinate $s_1$ in (\ref{eq:xsect3d_phi}) and the field line coordinate $t$ are connected by a linear equation, and one can replace the other in the remaining 2 equations from combining (\ref{eq:px_linear}) and (\ref{eq:xsect3d}).
This bilinear system can then be reduced to a quadratic equation for one coordinate, and the remaining ones can be obtained after back substitution.
If the intersection check succeeds, the surface coordinate tuple $\vec{u} \, = \, (i + s_1, \, j + s_2)$ is returned.

\section{Analysis} \label{sec:analysis}

The analysis module provides tools which can be applied regardless of the selected implementation of the magnetic field and boundary geometry (although some distinction between tokamak and stellarator configuration remains).
The tools are integrated into a collection of frontend tasks that store results in output files for visualization.
At the core of this is tracing of magnetic field lines (section \ref{sec:fieldline_tracing}), and this integrated into the construction of \Poincare maps (section \ref{sec:poincare_maps}) and invariant manifolds of X-points (section \ref{sec:invariant_manifolds}).
The workhorse application is the computation of the field line connection length for sets of initial points (section \ref{sec:connection_length}).
Supporting analysis tools (section \ref{sec:misc_tasks}) for the construction of flux surfaces and for the Fourier transform of perturbation fields are provided.

\subsection{Field line tracing} \label{sec:fieldline_tracing}

Magnetic field lines are curves in space which follow the direction of the magnetic field $\vec{B}(\vec{r})$ at each point along its length.
Field lines can be constructed by starting from an initial point $\vec{p} \, = \, (r_0, z_0, \varphi_0)$ and integrating a system of ordinary differential equations.
In toroidal plasma confinement systems, the angle $\varphi$ can serve as an independent time-like variable.
Then, $\vec{F}_{\vec{p}}(\varphi) \, = \, (r, z)$ is a field line through $\vec{p}$ if it is a solution of

\begin{equation}
\frac{dr}{d\varphi} = \frac{r \, B_r}{B_\varphi}, \qquad
\frac{dz}{d\varphi} = \frac{r \, B_z}{B_\varphi}. \label{eq:fieldline}
\end{equation}

\begin{figure}
\centering
\includegraphics[width=65mm]{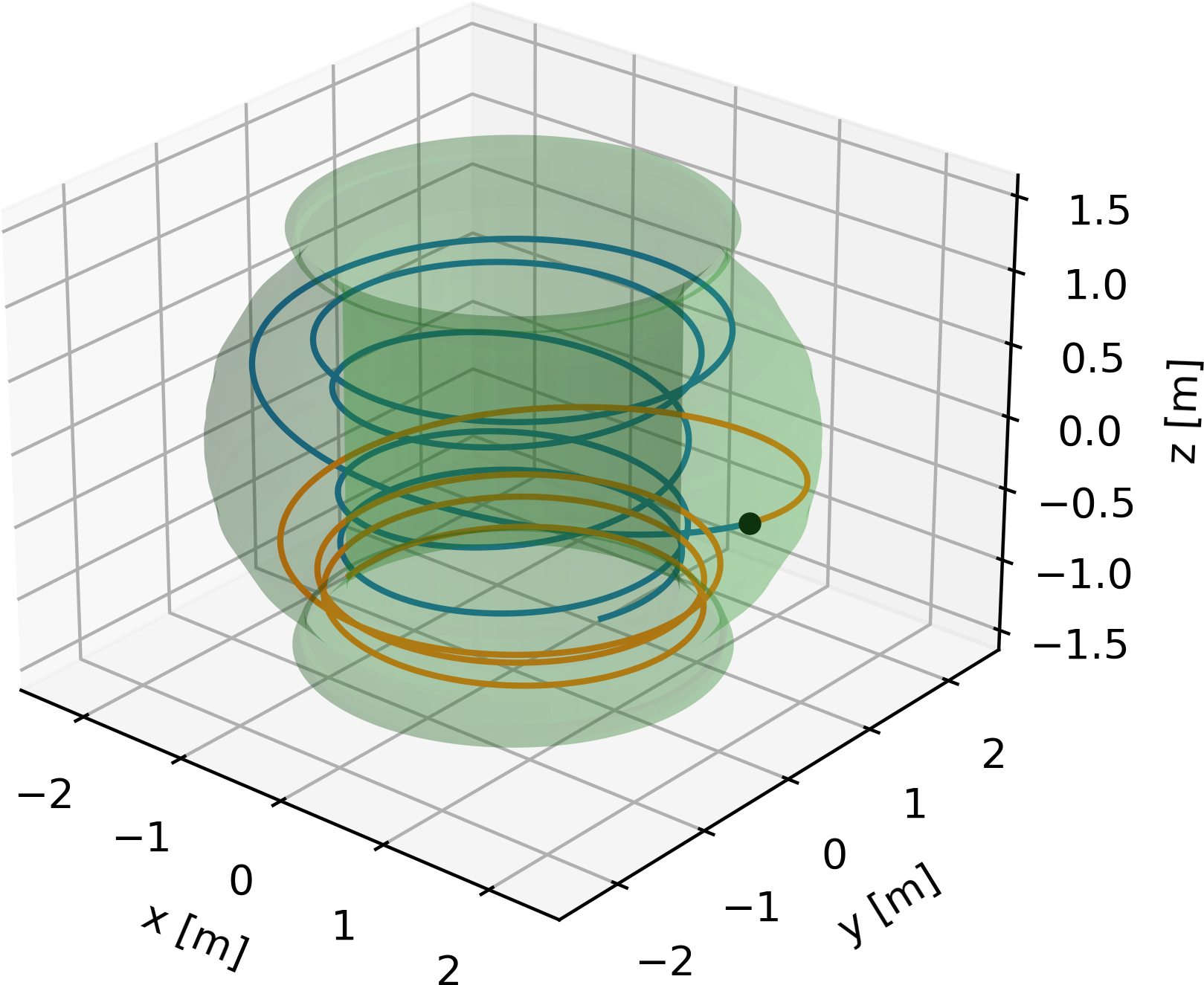}
\caption{Field line in the scrape-off layer (SOL) of KSTAR. The initial point $\vec{p}$ is marked by the black dot. Forward (blue) and backward (orange) traces are shown. Tracing is terminated on material surfaces.}
\label{fig:fieldline}
\end{figure}

with $\vec{F}_{\vec{p}}(\varphi_0) \, = \, (r_0, z_0)$.
Integration of (\ref{eq:fieldline}) can be done in either clockwise (decreasing $\varphi$) or counter-clockwise (increasing $\varphi$) direction.
An example is shown in figure \ref{fig:fieldline}.
Numerical methods have been established for solving ordinary differential equation initial value problems \cite{Hairer2000, NumericalRecipes2007}.
A family of embedded Runge-Kutta methods \cite{Dormand1980, Prince1981, Dormand1986} is implemented for solving (\ref{eq:fieldline}), and adaptive step-size adjustment is applied for error control.
The 5th order method by Dormand and Prince (\texttt{dopr5}) is a good general-purpose integrator and is used by default.
Furthermore, two variable-coefficient linear multi-step methods in Nordsieck form are implemented through an interface with LSODE \cite{Hindmarsh1983, LSODE} (see \ref{sec:fieldline_accuracy}).

Field line tracing is terminated on material surfaces.
A small tolerance $\varepsilon \sim 10^{-5} \, \textendash \, 10^{-7} \, \meter$ is required for the local error of the integration step in order to avoid significant error accumulation over the entire field line path (see figure \ref{fig:accuracy_benchmark}).
For the intersection point on the boundary, on the other hand, a much higher tolerance of $\varepsilon_{\textnormal{xsect}} \, = \, 0.1 \, \milli\meter$ to $1 \, \milli\meter$ is sufficient for most applications.
A cubic Hermite spline approximation of the field line segment comes at no additional cost, because the right hand sides of (\ref{eq:fieldline}) need to be evaluated at $\vec{p}_{n}$ and $\vec{p}_{n+1}$ anyway for the $n$-th integration step (and possibly in preparation of the next one).
For $\varepsilon_{\textnormal{xsect}} \, \gg \, \varepsilon$, this should be close enough to the exact solution.
In particular, dense output consistent with the order of the integration method - and the additional $\vec{B}(\vec{r})$ evaluations that come along - should not be required.

\begin{figure}
\centering
\includegraphics[width=55mm]{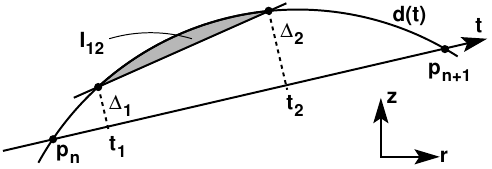}
\caption{Cubic Hermite spline $d(t)$ representation of field line segment $\vec{p}_{n} \rightarrow \vec{p}_{n+1}$. A piecewise linear approximation between $t_1$ and $t_2$ is shown with $\Delta_1 = d(t_1)$ and $\Delta_2 = d(t_2)$.}
\label{fig:xsect}
\end{figure}

The intersection check (\ref{eq:xsect}) is based on a linear approximation of a field line segment.
Let $d(t)$ with $t \in [0,1]$ be a cubic Hermite spline that represents the distance between the linearized segment and the {\it exact} solution as shown in figure \ref{fig:xsect}.
Then

\begin{equation}
I(1) \, = \, \frac{|d_{0}' \, - \, d_{1}'|}{12}
\end{equation}

is the average error of the linearization based on the error integral

\begin{equation}
I(t) \, = \, \int_0^t d\tilde{t} \, \left| d(\tilde{t}) \right|.
\end{equation}

For $I(1) > \varepsilon_{\textnormal{xsect}}$, 
one can iteratively construct piecewise linear approximations until the required accuracy is achieved.
E.g. for the sub-segment $[t_1, t_2]$ shown in figure \ref{fig:xsect}, the integral error is

\begin{equation}
I_{12} \, = \, I(t_2) \, - \, I(t_1) \, - \, \frac{\Delta_1 \, + \, \Delta_2}{2} \, \left(t_2 \, - \, t_1\right).
\end{equation}

For $\varepsilon_{\textnormal{xsect}} \, = \, 0.1 \, \milli\meter$, less than two sub-segments are required on average for the case study in figure \ref{fig:Bevaluations} with $\varepsilon \, = \, 10^{-7} \, \meter$.
However, an average of up to 4 sub-segments can be necessary for the larger integration steps that are sufficient for the higher order (\texttt{dopr8}) method at $\varepsilon \, = \, 10^{-5} \, \meter$.

\subsection{\Poincare maps} \label{sec:poincare_maps}

\begin{figure}
\centering
\includegraphics[width=55mm]{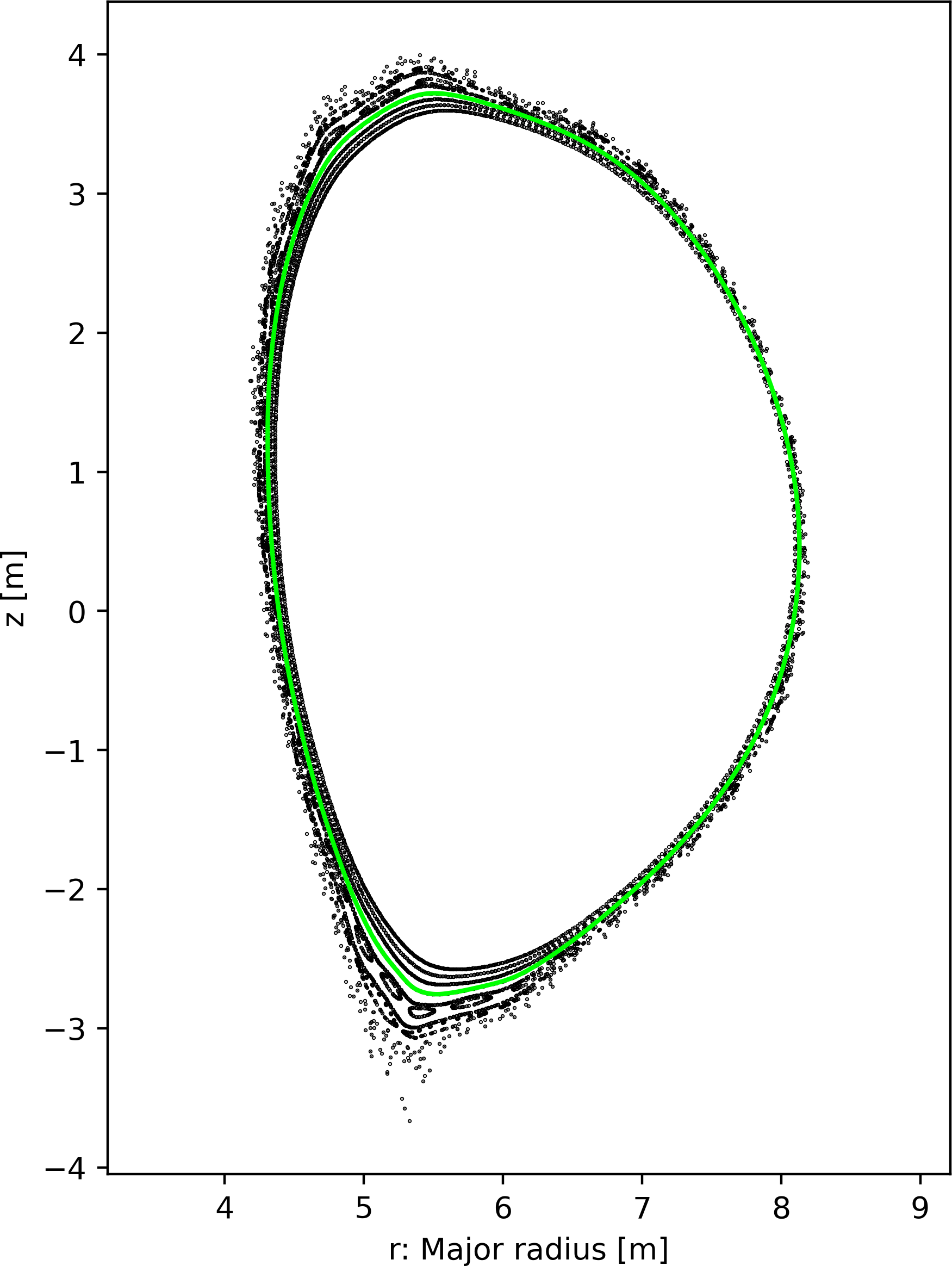}
\caption{\Poincare map for ITER with $n = 3$ RMPs \cite{Frerichs2024} including plasma response from MARS-F.
From inside to outside, one can see closed flux surfaces, magnetic island chains, and a chaotic region from overlap of neighboring island chains.
A B-spline curve $\vec{P}_{\textnormal{fit}}(\theta)$ (green) is fitted to one set of points (see section \ref{sec:fluxsurf3d}).}
\label{fig:poincare_map}
\end{figure}

\begin{figure}
\centering
\includegraphics[width=80mm]{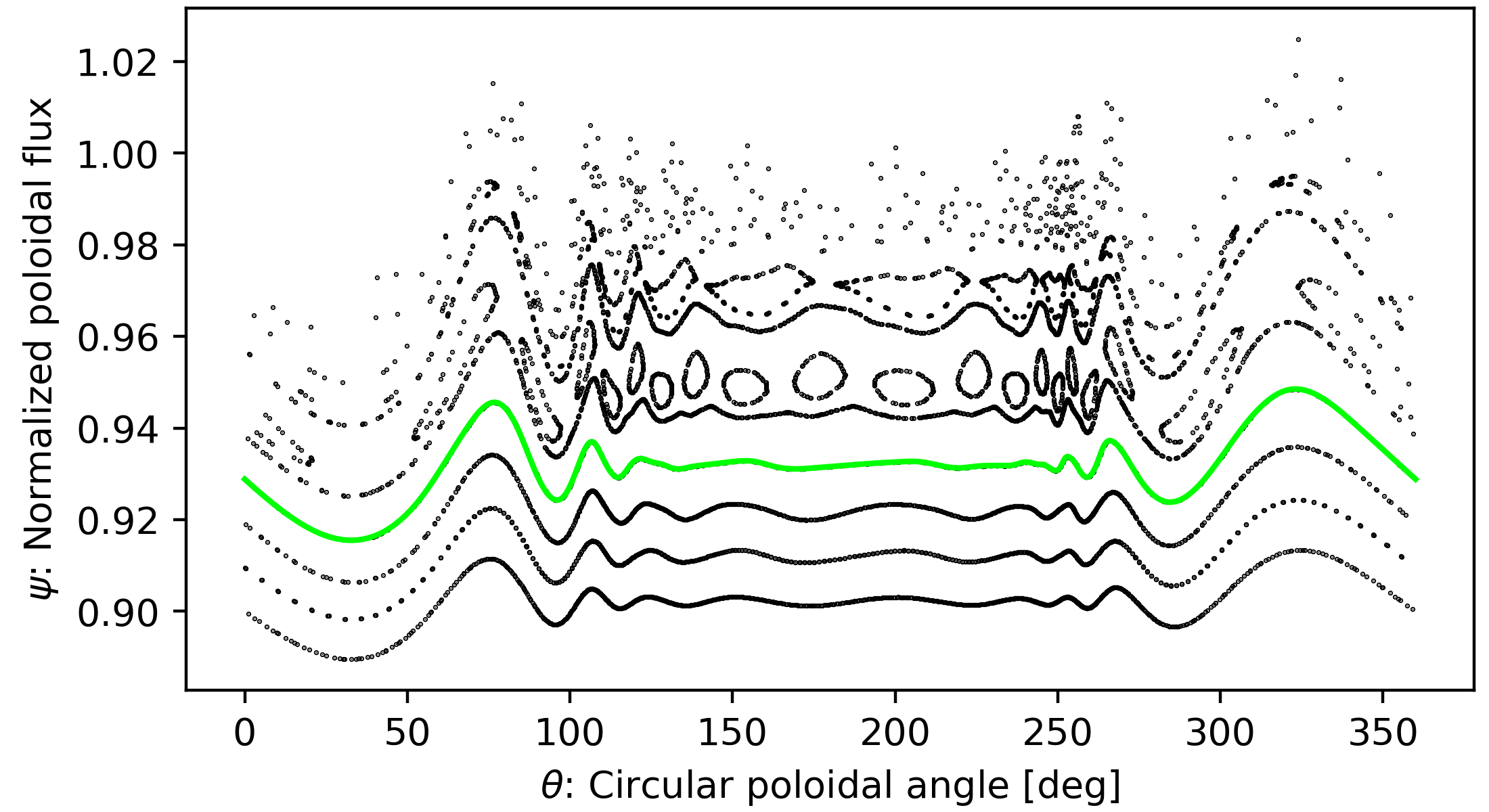}
\caption{Same case as figure \ref{fig:poincare_map}, but in $(\theta, \psiN)$ coordinates.
A B-spline function $\psiN_{\textnormal{fit}}(\theta)$ (green) is fitted to $\psiN(\vec{p}_i)$ values.
}
\label{fig:poincare_map_theta-psiN}
\end{figure}

Given the toroidal nature of magnetic confinement systems, \Poincare maps are a useful tool for evaluating properties of magnetic field lines.
A \Poincare map is defined by the return points of periodic orbits of a continuous dynamical system with a lower-dimensional transversal subspace, the \Poincare section.
For magnetic field lines, the R-Z plane at a selected angle $\varphi_S$ is a suitable location for a \Poincare section $S$.
The \Poincare map for systems with toroidal symmetry \nsym is then given by

\begin{equation}
\vec{P}_{S}^{\pm}(r,z) \, = \, \vec{F}_{(r,z,\varphi_S)}\left(\varphi_S \, \pm \frac{2 \, \pi}{\nsym}\right), \label{eq:poincare_map}
\end{equation}

i.e. it maps the point $(r,z,\varphi_S)$ to its position along the field line after one field period.
Visualization of \Poincare maps is based on the iterative application $\vec{p}_{i+1} \, = \, \vec{P}_S(\vec{p}_{i})$ for one or several starting points.
An example is shown in figure \ref{fig:poincare_map}.
From inside to outside, one can see that field lines form closed flux surfaces, magnetic island chains, and a chaotic region from overlap of neighboring island chains.
The same case is shown in figure \ref{fig:poincare_map_theta-psiN}, but now the circular poloidal angle $\theta$ and the normalized poloidal flux \psiN are used as coordinates instead of $(r,z)$.
The latter figure is more suitable for the analysis of the plasma edge region, but this is - pending an implementation of a \psiN-like radial coordinate for stellarators - only available for tokamak configurations for now.

\subsection{Invariant manifolds} \label{sec:invariant_manifolds}

The locations where material surfaces are most impacted by the plasma are determined by the strike points of the (primary) magnetic separatrix.
The magnetic separatrix - as well as any other flux surface - is an invariant manifold, i.e. any point on it remains within the manifold while moving along a field line.
However, the invariant manifolds associated with an X-point (hyperbolic point) are susceptible to small magnetic perturbations, and they split into two distinct sets that intersect each other in so called homoclinic tangles \cite{Evans2004a}.
They appear as helical lobes around poloidally diverted plasmas and form the boundary for perturbed field lines that connect from the plasma interior to divertor targets and bring heat and particles along with them.
In literature on non-linear dynamics (e.g. \cite[p. 45]{Guckenheimer1983}, \cite[p. 560]{Lichtenberg1992} or \cite{Wiggins2003}), invariant manifolds are often referred to as either stable or unstable, depending on if they approach or move away from the hyperbolic point.
These labels are, however, less suitable for toroidal systems where field lines can be traced in either direction of the time-like variable $\varphi$.
Let $\vec{X} = (r_x, z_x)$ be the X-point of an axisymmetric equilibrium, then the invariant manifolds associated with it are defined as

\begin{figure}
\centering
\includegraphics[width=70mm]{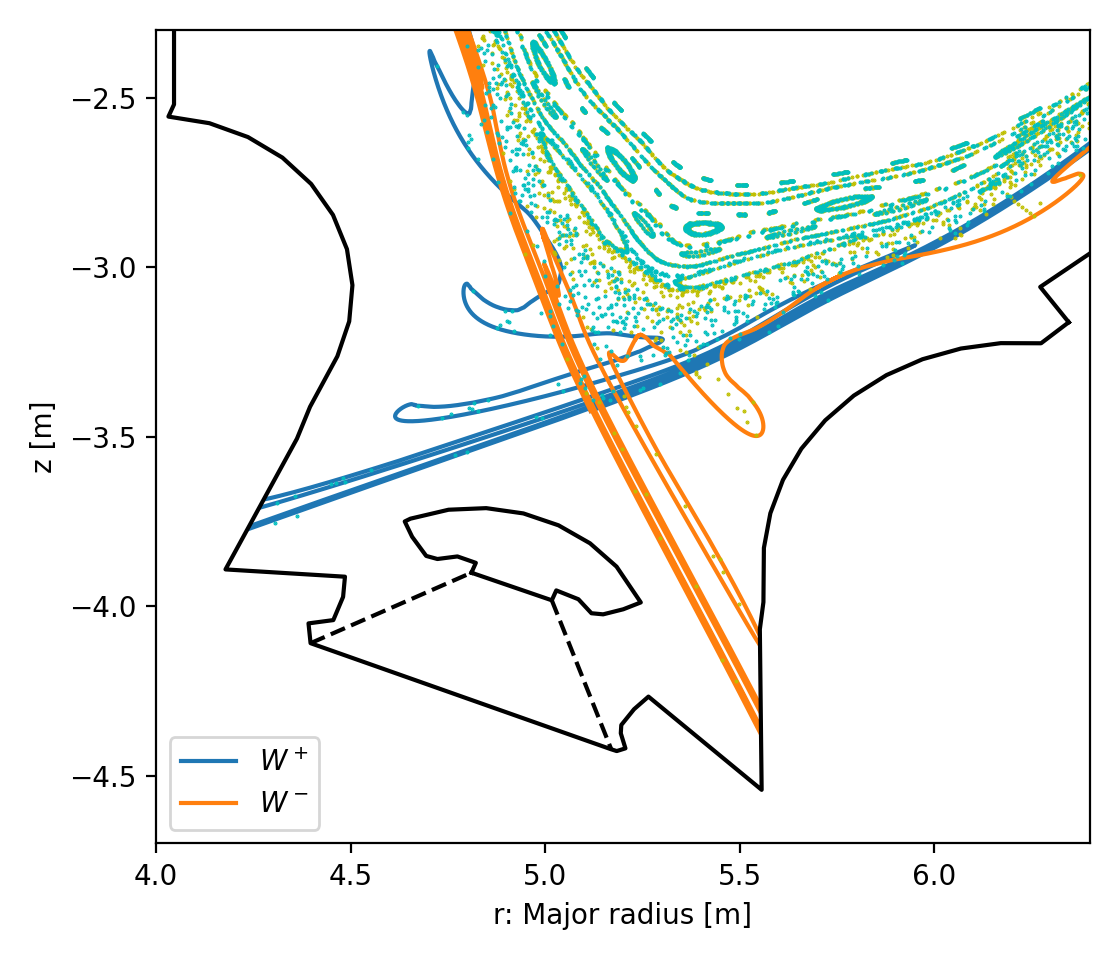}
\caption{Invariant manifolds for the same ITER case with RMPs as in figure \ref{fig:poincare_map}. Higher resolution \Poincare maps from forward (yellow) and backward (cyan) tracing are shown.}
\label{fig:invariant_manifolds}
\end{figure}

\begin{equation}
W_\vec{X}^{\pm} \, = \, \left\{\vec{p} \Big| \lim_{\varphi \, \rightarrow \, \pm\infty} \, \vec{F}_{\vec{p}}(\varphi) \, \rightarrow \, \vec{X}\right\}.
\end{equation}

Of particular interest are the sets with field lines that approach the X-point from the plasma side, which are the ones that form the last closed flux surface in configurations without perturbations.
Computation of those is implemented as follows.
First, starting points are constructed by moving a small step $\varepsilon_X$ off the X-point in direction of the eigenvectors of the field line Jacobian.
Then, starting points are distributed toroidally within, and field lines are traced in the opposite direction in which they would asymptotically approach the X-point.

An example is shown in figure \ref{fig:invariant_manifolds}.
It can be seen that the invariant manifolds oscillate wildly around the X-point and intersect the divertor targets.
The additional \Poincare maps show how field lines are guided by those helical lobes structures to the divertor targets.

\subsection{Connection length} \label{sec:connection_length}

\begin{figure}
\centering
\includegraphics[width=70mm]{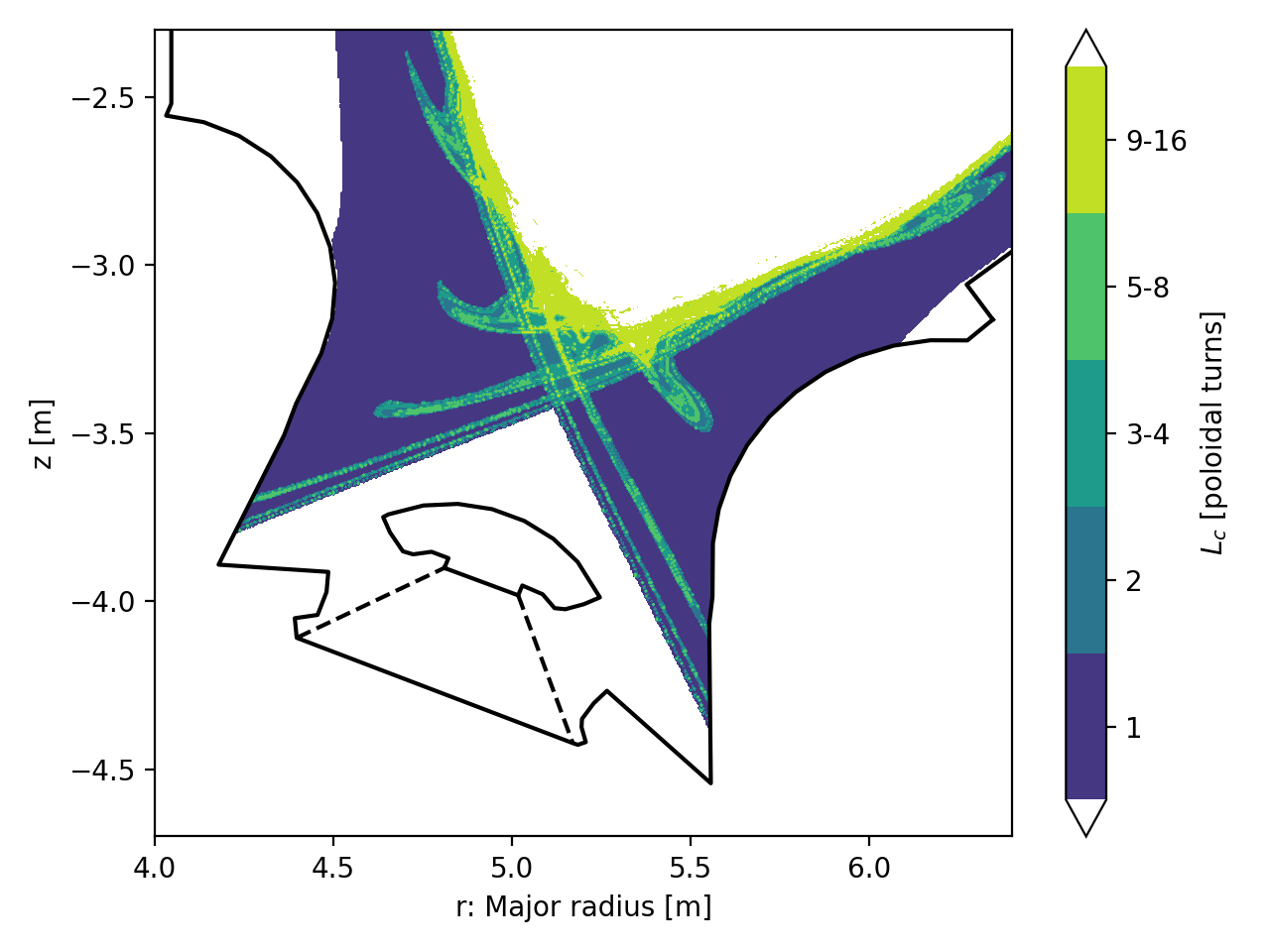}
\caption{Field line connection length for the same ITER case with RMPs as in figures \ref{fig:poincare_map} and \ref{fig:invariant_manifolds}. Field line tracing is terminated after $1000 \, \meter$ in either direction. Scrape-off layer field lines connection from inner to outer divertor target within one poloidal turn. The helical lobes of the magnetic separatrix (invariant manifolds) guide field lines from the interior to the divertor targets in two or more poloidal turns.}
\label{fig:connection_length}
\end{figure}

\begin{figure}
\centering
\includegraphics[width=70mm]{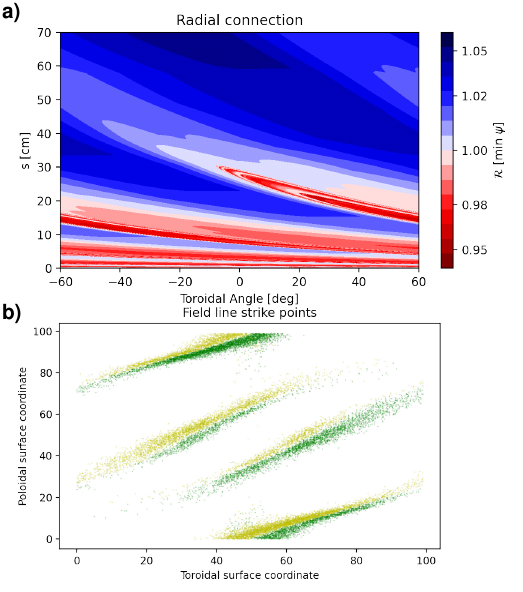}
\caption{Magnetic footprints: (a) radial connection of magnetic field lines on the outer divertor target in ITER for the same case as in figure \ref{fig:connection_length}. Particle and heat loads are expected where field lines connection into the plasma ($\mathcal{R} < 1$). Characteristics are $s_{\textnormal{max}} = 30.6 \, \centi\meter$, $A = 1.677 \, \meter^2$ (for $120 \, \deg$), $\overline{\mathcal{R}} = 0.987$ and $\min\mathcal{R} = 0.958$.
(b) Strike points on the boundary of Wistell-A \cite{Bader2020} from field line diffusion. Green and yellow colors indicate strike points in forward and backward direction, respectively.}
\label{fig:footprints}
\end{figure}

\Poincare maps give useful information where field lines are confined for many toroidal turns, and invariant manifolds show how field lines are guided to divertor targets.
Complementary figures of merit for {\it open} field lines are the connection length $L_c$, i.e. the distance along the field line from one target to the other, and the radial connection $\mathcal{R}$ which depicts from how far inside the plasma a perturbed field line connects.
This is typically computed for a large number $N$ of starting points, e.g. in the R-Z plane at some toroidal location as shown in figure \ref{fig:connection_length}, or on the divertor targets as shown in figure \ref{fig:footprints} (a).

The spatial layout of starting points $\vec{p}_i$ is irrelevant for the loop $i \, = \, 1, \ldots, N$, and a number of different grid types are implemented for visualization \cite{MOOSE}.
Supporting procedures for grid construction are provided (\texttt{rzmesh}, \texttt{footprint\_grid}).
For each starting point $\vec{p}_i$, field lines are traced in both directions until they intersect material surfaces or some cut-off value $L_{c, \textnormal{max}}$ is reached.
The connection length is stored in both arc length and poloidal turns (the latter of which is shown in figure \ref{fig:connection_length}), and the radial connection is computed from the minimum of \psiN values along each field line.
The magnetic footprint in figure \ref{fig:footprints} (a) shows where perturbed field lines are expected to bring particle and heat loads to the divertor target.

While the radial connection $\mathcal{R}$ is a useful figure of merit for configurations where an equilibrium is perturbed, a different approach is necessary for stellarators.
Field line diffusion can be used as proxy for particle and heat loads \cite{Punjabi2014, Bader2017}.
For this, a grid of starting points $\vec{p}_i$ is first generated on a good flux surface near the boundary.
Then, field lines are traced from each $\vec{p}_i$ - but after each integration step with arc length $s$, a displacement $\Delta_{rz} \, = \, \sqrt{4 \, \mathcal{D} \, s}$ is added in random direction within the R-Z plane.
Finally, intersection points on the boundary are plotted.
An example is shown in figure \ref{fig:footprints} (b) using surface coordinates (see section \ref{sec:boundary}).
Strike points can be binned (\texttt{strike\_point\_density}) in order to construct a proxy for particle and heat loads. 
The field line diffusion coefficient can be chosen as $\mathcal{D} \, = \, D / u$ to mimic particle transport with cross-field diffusivity $D$ and velocity $u$ along field lines.
Field line diffusion is beyond the originally intended application of FLARE, and it is better suited for integration into plasma boundary codes (e.g. EMC3-lite \cite{Feng2022}).
Nevertheless, it can be useful for scoping studies to aid and verify magnetic mesh generation (section \ref{sec:mmesh}).

\subsection{Supporting procedures} \label{sec:misc_tasks}

Miscellaneous procedures are provided in support of analysis tasks.
For tokamaks, equilibrium flux surfaces can be constructed from iso-\psiN contours (\texttt{equi2d\_contour}, \texttt{equi2d\_separatrix}).
This supports definition of the straight field line poloidal angle $\vartheta$ (\texttt{equi2d\_poloidal\_angle}) and computation of the Fourier transform of perturbation fields.
For any non-axisymmetric configuration, 3D flux surfaces can be constructed from B-spline fits to \Poincare maps.
Grids can be generated from either of those (\texttt{fluxsurf2d\_grid} and \texttt{fluxsurf3d\_grid}).

\subsubsection{Fourier transform} \label{sec:fourier_transform}

Spectral analysis of magnetic perturbations \dB in tokamaks is based on the magnetic flux

\begin{equation}
\Phi \, = \, \mathcal{J} \, \dB \cdot \nabla \psi
\end{equation}

perpendicular to the equilibrium flux surfaces. Here, $\mathcal{J} \, = \, \left(\vec{B} \cdot \nabla\vartheta\right)^{-1}$ is the Jacobian of the equilibrium magnetic coordinates $(\psi, \vartheta, \varphi)$.
The Fourier harmonics of the flux are

\begin{equation}
\Phi_{mn}(\psi) \, = \, \frac{1}{\left( 2 \pi \right)^2} \, \oint \!d\vartheta d\varphi \, \Phi \, e^{-i \left(m \vartheta \, - \, n \varphi\right)}. \label{eq:Phimn}
\end{equation}

The resonant harmonics $m \, = \, n \, q(\psi)$ are independent of the magnetic coordinate system, and they determine the width of magnetic islands \cite{Park2008}.
This can be normalized with respect to the poloidal flux:

\begin{equation}
\Phi_{mn}^\ast \, = \, \frac{\Phi_{mn}}{\psisepx \, - \, \psiaxis}.
\end{equation}

\begin{figure}
\centering
\includegraphics[width=65mm]{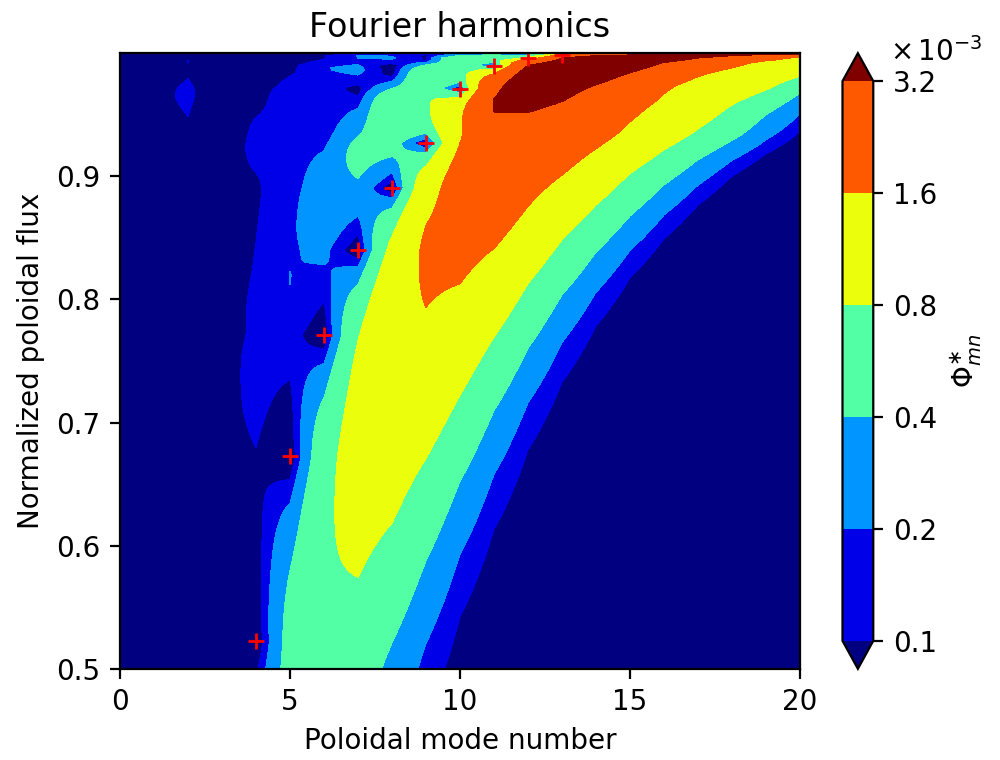}
\caption{Normalized Fourier harmonics $\Phi_{mn}^{\ast}$ for the same $n = 3$ RMP scenario in ITER as before. The position of resonances are marked by symbols.}
\label{fig:PhiNmn}
\end{figure}

An example is shown in figure \ref{fig:PhiNmn} for the same $n = 3$ RMP scenario in ITER as before.
It can be seen that the resonant harmonics are substantially screened which is due to plasma response.
At the same time, large non-resonant components are present which lead to helical displacements (kinks) of flux surfaces.

\subsubsection{Flux surfaces} \label{sec:fluxsurf3d}

Non-axisymmetric flux surfaces can be constructed from B-spline fits to \Poincare maps.
A field line is traced for many field periods, and \Poincare maps are generated at several toroidal locations along the way.
At each location, a B-spline approximation

\begin{equation}
\vec{P}_{\textnormal{fit}}(\theta) \, = \, \sum_{j=1}^{m} \, \vec{c}_j \, B_j^{(k)}(\theta) \label{eq:Pfit}
\end{equation}

is constructed with $m$ basis functions $B_j^{(k)}$ of order $k$ ($k = 4$ for cubic polynomials is set as default).
The B-spline coefficients $\vec{c}_j$ are determined from minimizing

\begin{equation}
\chi^2 \, = \, \sum_{i = 1}^{n} \, \left(\vec{P}_{\textnormal{fit}}(\theta_i) \, - \, \vec{p}_i\right)^2.
\end{equation}

for a sequence of return points $\vec{p}_i, i = 1,...,n$ that form a closed flux surface contour.
Internal knots along $\theta$ can be placed such that (on average) an equal number of points $\vec{p}_i$ are covered.
Similarly, a B-spline function $\psiN_{\textnormal{fit}}(\theta)$ can be fitted to $\psiN(\vec{p}_i)$ values.
Example are shown in figures \ref{fig:poincare_map} and \ref{fig:poincare_map_theta-psiN}.





\section{Magnetic mesh} \label{sec:mmesh}

\def\phimap{\ensuremath{\varphi_{\textnormal{map}}}\xspace}
In 3D plasma boundary modeling, a magnetic mesh facilitates separation of the fast transport along magnetic field lines from the much slower transport in cross-field direction.
A self-consistent solution of the edge plasma and neutral gas requires many iterations on the same magnetic field (i.e. hydrodynamics in a magnetic field approach).
A significant advantage for particle based methods is that a magnetic mesh supports fast reconstruction of field line segments.
As such, a magnetic mesh is fundamental for the Monte Carlo fluid code EMC3 \cite{Feng2000, Feng2005} which requires repeated evaluation of similar field line segments.
A brief introduction of field line reconstruction is given in section \ref{sec:mmesh_reconstruction}, followed by a description of the magnetic mesh generator in section \ref{sec:mmesh_generator}.

\subsection{Field line reconstruction} \label{sec:mmesh_reconstruction}

\begin{figure}
\centering
\includegraphics[width=80mm]{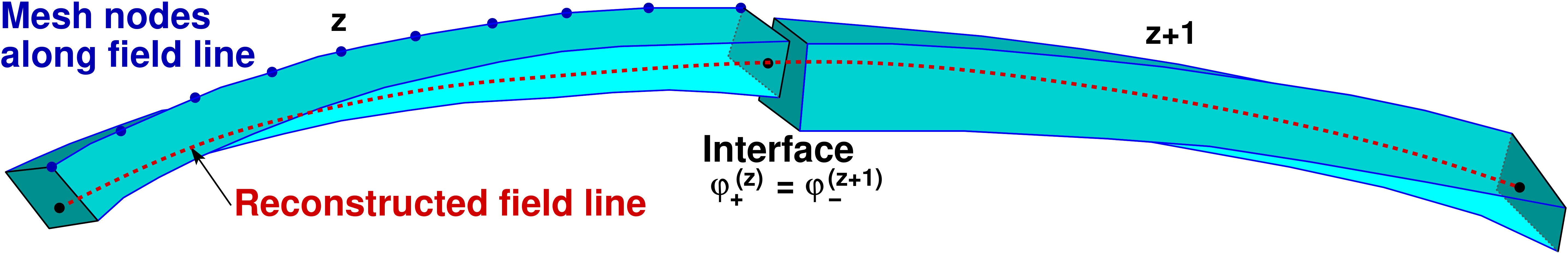}
\caption{Interpolation of a field line within two adjacent flux tubes with a common interface $\varphi_+^{(z)} \, = \, \varphi_-^{(z+1)}$. The cross-section of the flux tubes can change along the toroidal direction due to magnetic shear and perturbations. Complete overlap is not required at the interface, but the quadrilateral cross-sections must be convex for a reversible mapping of the local coordinates $(\xi, \eta)$.}
\label{fig:finite_flux_tubes}
\end{figure}

Field line reconstruction is based on the interpolation of field line segments within a finite flux tube and a mapping of local coordinates from one finite flux tube to the next.
Figure \ref{fig:finite_flux_tubes} shows two adjacent finite flux tubes and a field line that runs through both of them.
Let $\vec{F}_i(\varphi), i = 1,\ldots,4$ be four field lines that form a flux tube $\vec{z}$ over the domain $T_{\vec{z}} \, = \, [\varphi_-, \varphi_+]$, then an interpolant field line segment is given by

\begin{equation}
\vec{F}_{\xi \eta}(\varphi) \, = \, \sum_{i \, = \, 1}^{4} \vec{F}_{i}(\varphi) \, N_{i}(\xi, \eta) \label{eq:interpolantF}
\end{equation}

at any location $\varphi \in T_{\vec{z}}$.
Within the flux tube cross-section, the local coordinates $(\xi, \eta) \, \in \, [-1,1]^2$ define the position of the interpolant field line through the shape functions

\begin{equation}
N_i(\xi, \eta) \, = \, \frac{1}{4} \, \left(1 \, + \, \xi_i \, \xi\right) \, \left(1 \, + \, \eta_i \, \eta\right)
\end{equation}

\begin{figure}
\centering
\includegraphics[width=50mm]{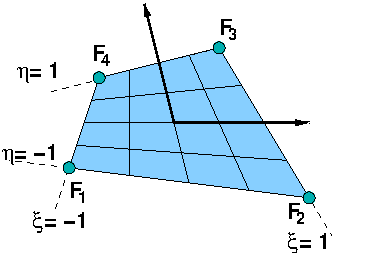}
\caption{Cross section of a flux tube formed by 4 field lines $\vec{F}_1 \, \ldots \, \vec{F}_4$ and local coordinate system $(\xi, \eta)$.}
\label{fig:mmesh_quadrilateral}
\end{figure}

where $(\xi_i, \eta_i)$ are the coordinates associated with the four $\vec{F}_i$ as defined in figure \ref{fig:mmesh_quadrilateral}.
Continuity at the interface \phimap between two adjacent flux tubes $\vec{z}$ and $\vec{z'}$ requires

\begin{equation}
\vec{F}_{\xi \eta}^{(\vec{z})}(\phimap) \, = \, \vec{F}_{\xi' \eta'}^{(\vec{z'})}(\phimap)
\end{equation}

where both left and right hand sides are evaluated according to (\ref{eq:interpolantF}) from a corresponding set of field line segments.
This implies that (\ref{eq:interpolantF}) needs to be inverted in order to map from one set of coordinates $(\vec{z}, \xi, \eta)$ to the other $(\vec{z'}, \xi', \eta')$, or vice versa.
An advantage of bilinear interpolation is that it allows a reversible mapping at the interface without iteration or approximation that non-linear problems typically imply.
A convex cross-section is sufficient for a unique inverse mapping of the local coordinates, but mapping of the flux tube itself does not need to be unique.
The mapping procedure in EMC3 is supplemented with an algorithm for finding the correct neighbor \cite{Frerichs2010}.

\subsection{Mesh generator} \label{sec:mmesh_generator}

The computational domain does not need to cover the full $360 \, \deg$ and can be adapted to the symmetry of the magnetic configuration.
Depending on magnetic shear and/or perturbations, the domain may need to be split into several blocks.
A block size of $40 \, \deg$ is often used for RMP configurations in tokamaks, and a block size of $36 \, \deg$ is used to cover half a field period in W7-X.
The strategy to construct a magnetic mesh is as follows: determine the inner boundary with the core plasma (section \ref{sec:mmesh_inner_boundary}), construct a 2D base mesh at selected toroidal positions (section \ref{sec:mmesh_base_mesh}), trace field line segments from there across the toroidal domain (section \ref{sec:mmesh_tracing}), extend the mesh for coupling with neutral particle transport (section \ref{sec:mmesh_n0}), and finally generate an approximation of the divertor target geometry (section \ref{sec:mmesh_targets}).
A detailed description of mesh parameters can be found in the FLARE user manual \cite{FLARE}.

\subsubsection{Inner boundary} \label{sec:mmesh_inner_boundary}

\begin{figure}
\centering
\includegraphics[width=70mm]{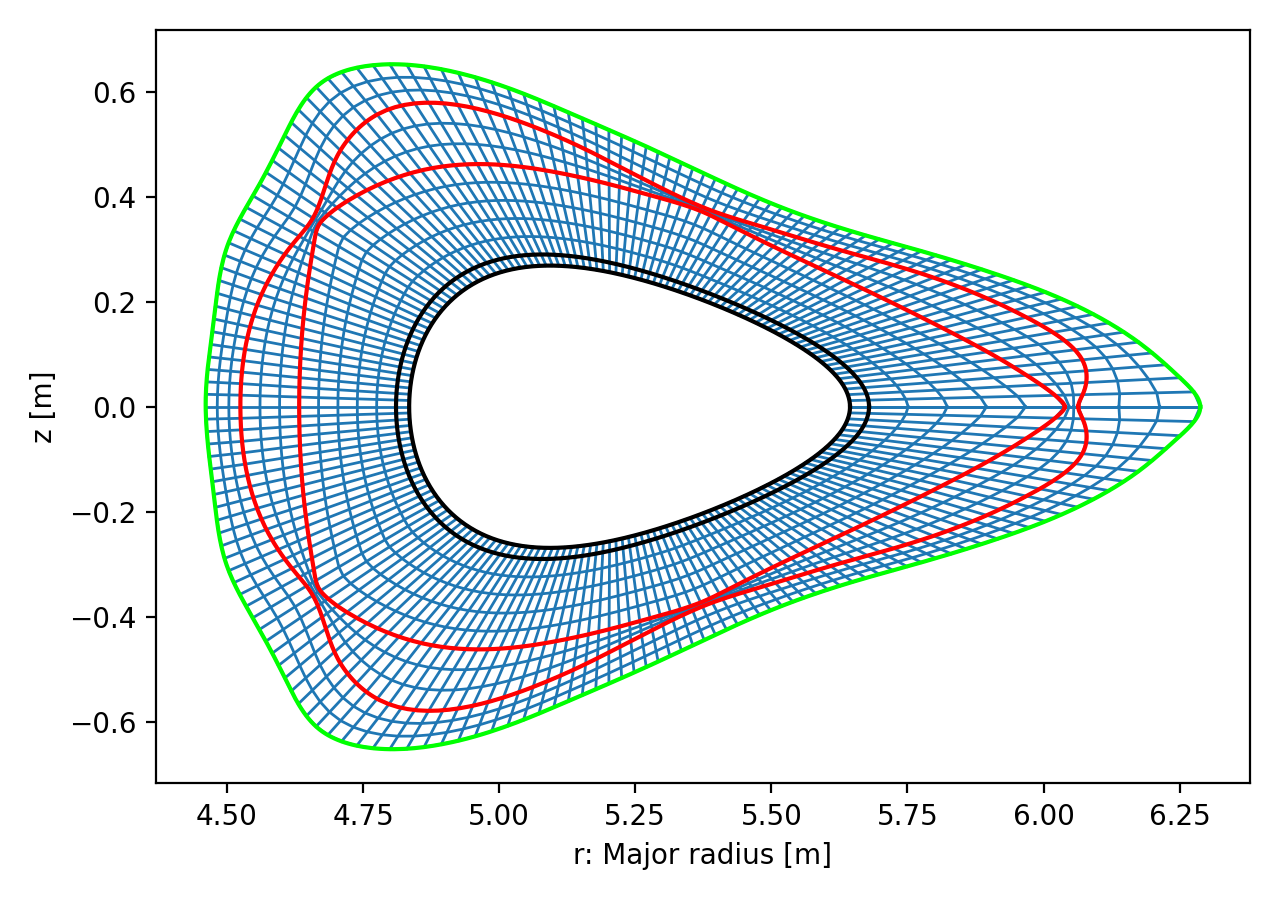}
\caption{Base mesh for W7-X standard divertor configuration. A coarse resolution is used for visualization purposes. Intermediate guiding contours (red) are applied around the 5/5 island chain.}
\label{fig:base_mesh}
\end{figure}

The inner simulation boundary must be a closed magnetic flux surface.
Furthermore, a second closed flux surface is required to construct cells at the inner mesh boundary.
This is useful for weighted sampling of sources at the inner boundary which takes into account steeper gradients where flux surfaces are closer together.
Suitable locations $\vec{p}_1$ and $\vec{p}_2$ can be found from \Poincare maps (see section \ref{sec:poincare_maps}).
Output of this step are B-spline approximations (\ref{eq:Pfit}) of the two flux surfaces (e.g. the black curves in figure \ref{fig:base_mesh}).

\subsubsection{Base mesh} \label{sec:mmesh_base_mesh}

Since the information of the magnetic field geometry is determined by the sequence of nodes along field lines segments, the mesh layout in the cross-field direction is free to choose.
Most stellarator applications (W7-X, HSX) use a regular structured mesh in the radial and poloidal direction.
An example is shown in figure \ref{fig:base_mesh}.
Intermediate guiding contours (red) can be used to align the mesh in certain regions, e.g. around the 5/5 island chain in W7-X.
The outer boundary (green) can be a flux surface (if available), or some user-defined closed curve.
All curves can be given as approximating B-spline, interpolating cubic spline or finite Fourier series.
The two former can be manipulated interactively with the \texttt{mcurve} program that is part of the FLARE package.
Non-equidistant node spacing in radial and poloidal direction is supported through user defined distribution functions that can be generated with the \texttt{mcdf} program.

\begin{figure}
\centering
\includegraphics[width=39mm]{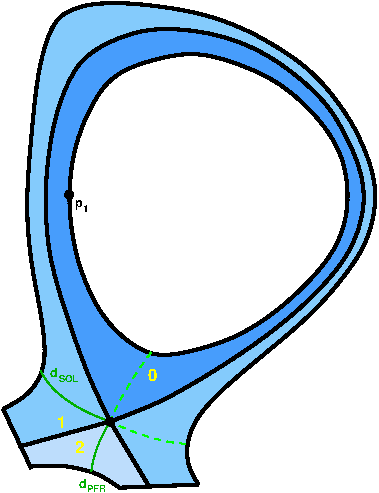}
\includegraphics[width=39mm]{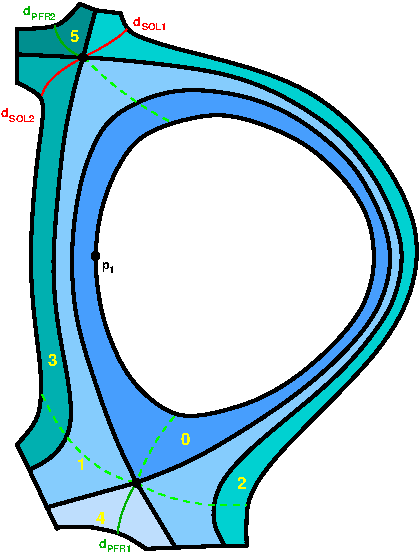}
\caption{Lower single null layout with 3 zones and disconnected double null layout with 6 zones. Within each zone, a regular structured mesh is used in the radial and poloidal direction. The zone layout and outer boundaries are equilibrium contours.}
\label{fig:mmesh_topo}
\end{figure}

\begin{figure}
\begin{center}
\includegraphics[width=80mm]{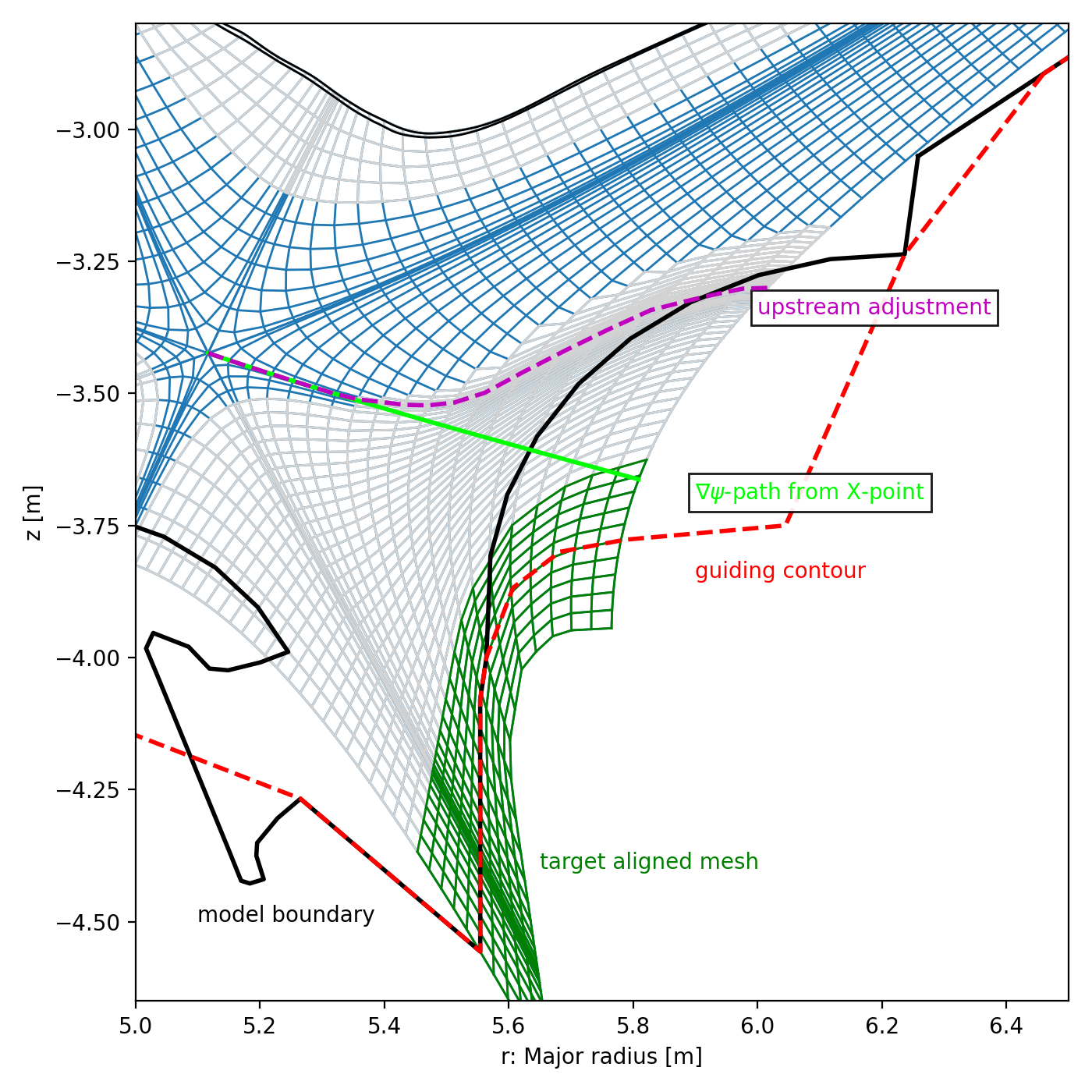}
\caption{Base mesh for ITER with subdomains for the main plasma, scrape-off layer (SOL) and private flux region (PFR). A quasi-orthogonal mesh is constructed upstream (blue), and a target aligned mesh is constructed on the downstream end (green) with interpolation along flux surfaces in between (gray).
This may require an upstream adjustment (purple) by pushing mesh nodes along flux surfaces.}
\label{fig:base_mesh_target}
\end{center}
\end{figure}

Subdomains are supported, and they are typically adapted to the equilibrium geometry in poloidal divertor tokamak applications \cite{Frerichs2010}.
The mesh generator can account for different magnetic field topologies such as single null and disconnected double null divertor configurations (see figure \ref{fig:mmesh_topo}).
A block-structured, quadrilateral base mesh is generated starting from the (main) equilibrium separatrix.
A quasi-orthogonal mesh is constructed upstream by tracing along the (equilibrium) $\nabla \psiN$ direction (blue mesh in figure \ref{fig:base_mesh_target}).
Configurations with RMPs required an intermediate layer of \texttt{n\_interpolate} interpolated flux surfaces (gray in figure \ref{fig:base_mesh_target}) between the perturbed inner boundary flux surface and the equilibrium aligned mesh.
The parameter \texttt{n\_interpolate} can be adapted as needed and must allow for enough room in case of strong non-resonant (kink) perturbations.

On the downstream end, the base mesh is aligned with the divertor targets (green in figure \ref{fig:base_mesh_target}).
For the base mesh located at $\varphi_b$, an aligned node $\vec{x}_{k}$ implies that a field line $\vec{F}_{\vec{x}_{k}} (\varphi)$ connects to the target at $\varphi_k$.
Typically, $\varphi_b$ is selected at the center of the toroidal domain, and mesh nodes are extended beyond the target in order to avoid gaps in the 3D mesh (see section \ref{sec:mmesh_tracing}).
The target alignment downstream in conjunction with the quasi-orthogonal mesh upstream requires an interpolated mesh in between (gray in figure \ref{fig:base_mesh_target}).
In the near SOL and PFR, the quasi-orthogonal mesh can be extended for \texttt{npXqo} cells below the X-point.
However, and adjustment is required for the far SOL where the $\nabla \psiN$-path from the X-point (light green in figure \ref{fig:base_mesh_target} intersects the boundary.
This can be achieved by a combination of ``moving the target'' by means of a guiding contour (red dashed line) and ``moving mesh nodes upstream`` (purple dashed line).
The latter can be controlled by a user-defined mapping

\def\smove{\ensuremath{s_{\textnormal{move}}}\xspace}
\def\spush{\ensuremath{s_{\textnormal{push}}}\xspace}
\begin{equation}
\rho \, \mapsto \, (\smove, \, \spush)
\end{equation}

which determines the arc length \smove that a mesh node at radial coordinate $\rho$ is moved along a flux surface from the light green line in figure \ref{fig:base_mesh_target} to the dashed purple line.
This requires that the mesh further upstream is pushed back, which occurs over an arc length of $\spush > \smove$ along the same flux surface.

\subsubsection{Tracing} \label{sec:mmesh_tracing}

For each node $\vec{x}_{ij}$ of the base mesh(es), the 3D mesh is constructed by tracing field line segments to successive toroidal locations $\varphi_k, k = 0, \ldots, n$.
The only constraint is that the cross-sections of the quadrilateral flux tubes remain convex (otherwise the simulation domain needs to be split into several blocks), which can be achieved more readily by starting from a quasi-orthogonal base mesh at the center of the toroidal domain.
This can be verified with the \texttt{mmesh check} program once mesh construction is finished.

\subsubsection{Extended domain for neutral particles} \label{sec:mmesh_n0}

\begin{figure}
\centering
\includegraphics[width=75mm]{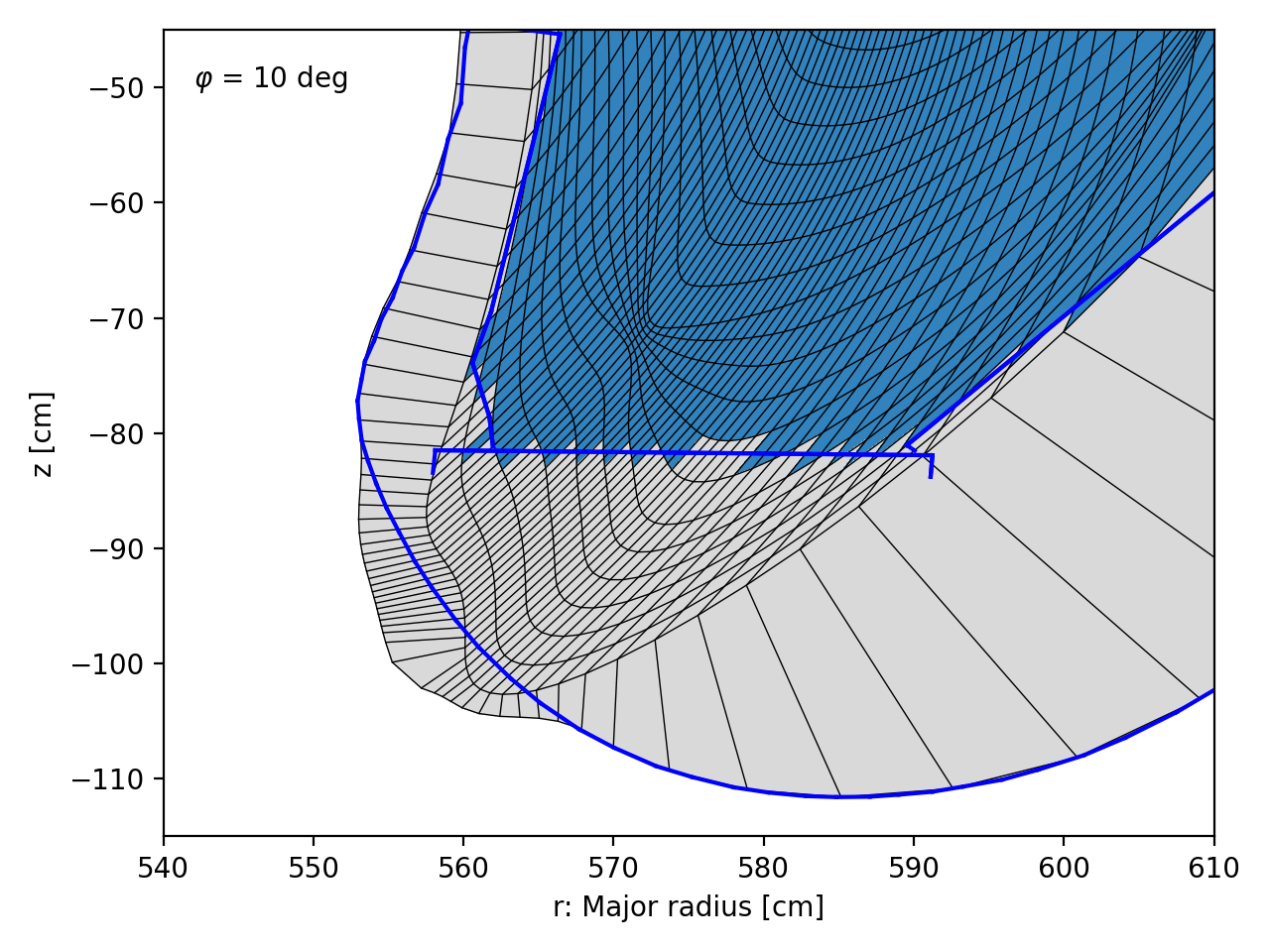}
\caption{R-Z slice through the 3D magnetic mesh for W7-X in standard divertor configuration. Plasma cells are highlighted in blue.
A coarse mesh is used for visualization purposes.
The \texttt{rzbuffer} option is used to automatically adjust the model boundary where needed.}
\label{fig:mmesh_n0+plates}
\end{figure}

The mesh can be extended beyond the plasma boundary in order to facilitate coupling with neutral particle transport (e.g. for particle balance control through pumping and gas fueling).
A number of different options are available for this tasks: 1) expand the plasma boundary by a given amount, 2) start from the model boundary (material surfaces) and automatically adjust where more space is needed (\texttt{rzbuffer}), or 3) take a user-defined surface.
An example for W7-X is shown in figure \ref{fig:mmesh_n0+plates}.

\subsubsection{Divertor targets} \label{sec:mmesh_targets}

The last step of the magnetic mesh construction process is generating an approximation of the boundary geometry by tagging cells that are out-of-bounds for the plasma.
This determines cell surfaces where boundary conditions for particle and energy transport are applied in EMC3.
The resulting fluxes are, however, mapped onto the actual boundary geometry for coupling with neutral particle transport and for post-processing.
Target alignment (available for tokamak configurations with axisymmetric boundaries) implies that a field line strikes within one cell length from the boundary cell surface.
Otherwise, a field line may require several cells in toroidal direction before it strikes the boundary, and adequate radial and poloidal resolution is necessary for oblique incident angles.

For poloidally closed boundaries, cells are marked as out-of-bounds if the cell center is outside of the boundary contour.
Some ambiguity remains with respect to cells that are partially out-of-bounds.
A refined out-of-bounds check considers a number of sub-volumes for each cell, and one can mark a cell as out-of-bounds either if at least one sample point is outside or if all sample points are outside.
An alternative for open boundaries (e.g. the divertor targets in W7-X) is to scan for intersections along the radial mesh direction.
An example is shown in figure \ref{fig:mmesh_n0+plates}.
The \texttt{mmesh rzslice} program can be used to verify the plasma domain within the magnetic mesh, and the \texttt{mmesh connection} program computes the field line connection length.


\section{Summary}

The FLARE code is a versatile tool set for the analysis of the magnetic geometry in non-axisymmetric tokamak and stellarator configurations in support of plasma boundary modeling and interpretation of experiments.
Interfaces for a number of different MHD equilibrium and plasma response models are implemented.
A magnetic mesh generator for fast reconstruction of field lines in 3D plasma boundary codes (EMC3-EIRNE) is included.


\appendix
\section*{Acknowledgements}

I would like to express my gratitude to the following colleagues for their support with interfacing plasma response models and with magnetic mesh generation for plasma boundary modeling.
In no particular order: Yuhe Feng (EMC3), Yueqiang Liu (MARS-F), Jong-Kyu Park (GPEC), Nate Ferraro (M3D-C${}^1$), John Schmitt (BMW), Yasuhiro Suzuki (HINT) and SangKyeun Kim (JOREK).

This work was supported by the U.S. Department of Energy under Awards DE-SC0012315, DE-SC0014210, DE-SC0020284 and DE-SC0020357, and by discretional funding from the College of Engineering at the University of Wisconsin-Madison.

\newpage
\section{Field line tracing accuracy} \label{sec:fieldline_accuracy}

\begin{figure}
\centering
\includegraphics[width=80mm]{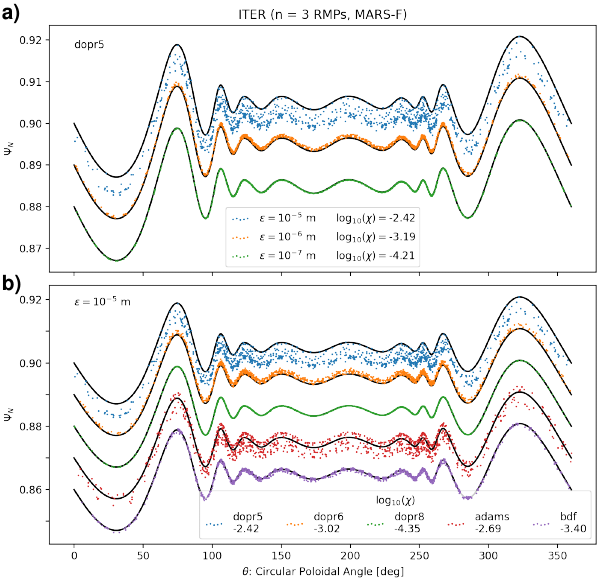}
\caption{\Poincare maps of the same perturbed flux surface for an ITER case \cite{Frerichs2020, Frerichs2021a} ($n = 3$ RMPs with MARS-F plasma response).
The vertical offset is introduced for visualization purposes.
a) Different error tolerances $\varepsilon$ for the \texttt{dopr5} method, and
b) different integration methods at $\varepsilon = 10^{-5} \, \meter$.
The black lines are from a B-spline fit to results for \texttt{dopr8} and $\varepsilon = 10^{-8} \, \meter$.}
\label{fig:accuracy_benchmark}
\end{figure}

The error tolerance parameter $\varepsilon$ for adaptive step size control during numerical integration of the field line path must be set to a low enough value in order to avoid error accumulation over many integration steps.
One way to evaluate a suitable $\varepsilon$ for field line tracing applications is by comparing \Poincare maps of a perturbed flux surface against a reference solution.
It is shown in figure \ref{fig:accuracy_benchmark} (a) that a 5th order embedded Runge-Kutta method by Dormand and Prince \cite{Dormand1980} (\texttt{dopr5}) produces visible deviations for $\varepsilon = 10^{-5} \, \meter$ and $10^{-6} \, \meter$, but reproduces the flux surface contour for $\varepsilon = 10^{-7} \, \meter$.
The overall accuracy is evaluated from

\begin{equation}
\chi \, = \, \sqrt{\frac{1}{N} \, \sum_{i = 1}^{N} \left(\psiN_{\textnormal{fit}}(\theta_i) \, - \, \psiN_i\right)^2} \label{eq:fiteval}
\end{equation}

where each of the \Poincare maps is constructed from $N = 1024$ return points.
The reference solution $\psiN_{\textnormal{fit}}$ is constructed from a B-spline fit to a \Poincare map obtained with an 8th order embedded Runge-Kutta method by Dormand and Prince \cite{Prince1981} (\texttt{dopr8}) and an error tolerance of $\varepsilon = 10^{-8} \, \meter$.
Figure \ref{fig:accuracy_benchmark} (b) shows that the \texttt{dopr8} method yields the same global accuracy already at $\varepsilon = 10^{-5} \, \meter$.
The other methods shown in figure \ref{fig:accuracy_benchmark} (b) are a 6th order embedded Runge-Kutta method by Dormand and Prince \cite{Dormand1986} (\texttt{dopr6}), and two variable-step, variable-order methods in Nordsieck form: an Adams-Moulton method of orders 1 to 12 (\texttt{adams}) and a backward differentiation formula method of orders 1 to 5 (\texttt{bdf}).
Both multi-step methods are implemented through LSODE \cite{Hindmarsh1983, LSODE}.

\begin{figure}
\centering
\includegraphics[width=80mm]{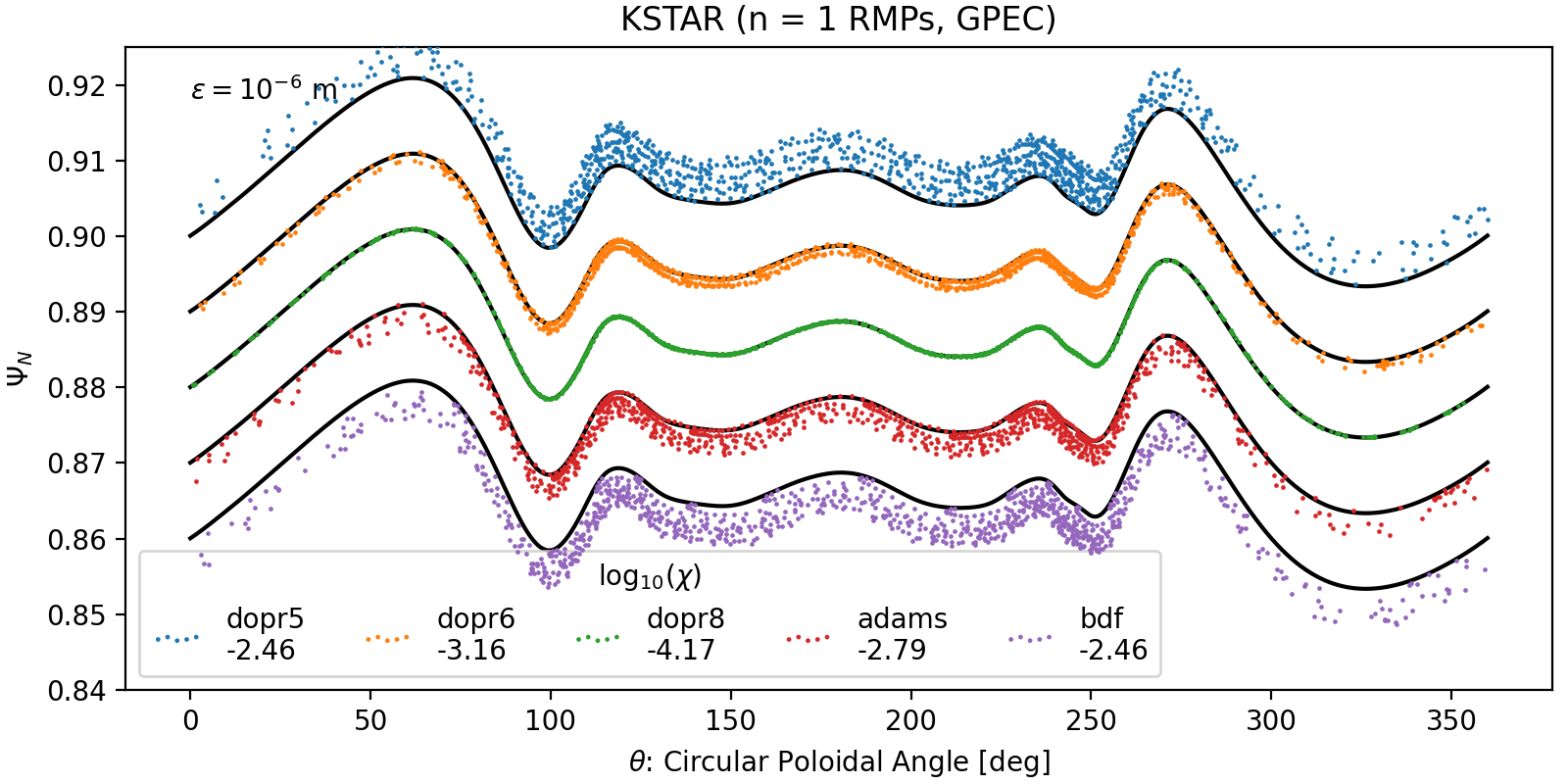}
\caption{\Poincare maps of the same perturbed flux surface for a KSTAR \cite{Frerichs2023} ($n = 1$ RMPs with GPEC plasma response) for different integration methods at $\varepsilon = 10^{-6} \, \meter$.
The vertical offset is introduced for visualization purposes.
The black lines are from a B-spline fit to results for \texttt{dopr8} and $\varepsilon = 10^{-8} \, \meter$.}
\label{fig:accuracy_benchmark_kstar}
\end{figure}

The \texttt{dopr8} method may get away with a more relaxed error tolerance - as long as one is able to identify an acceptable error level.
For field lines in chaotic regions, this may not be possible to evaluate, and even for perturbed flux surfaces it can depend on the application.
Figure \ref{fig:accuracy_benchmark_kstar} shows that comparable global accuracies require at least $\varepsilon = 10^{-6} \, \meter$ for the KSTAR case with $n = 1$ RMPs.
In particular, the multi-step methods achieve good global accuracy at $\varepsilon = 10^{-6} \, \meter$ for the ITER case but require $\varepsilon = 10^{-8} \, \meter$ for the KSTAR case.

\begin{figure}
\centering
\includegraphics[width=75mm]{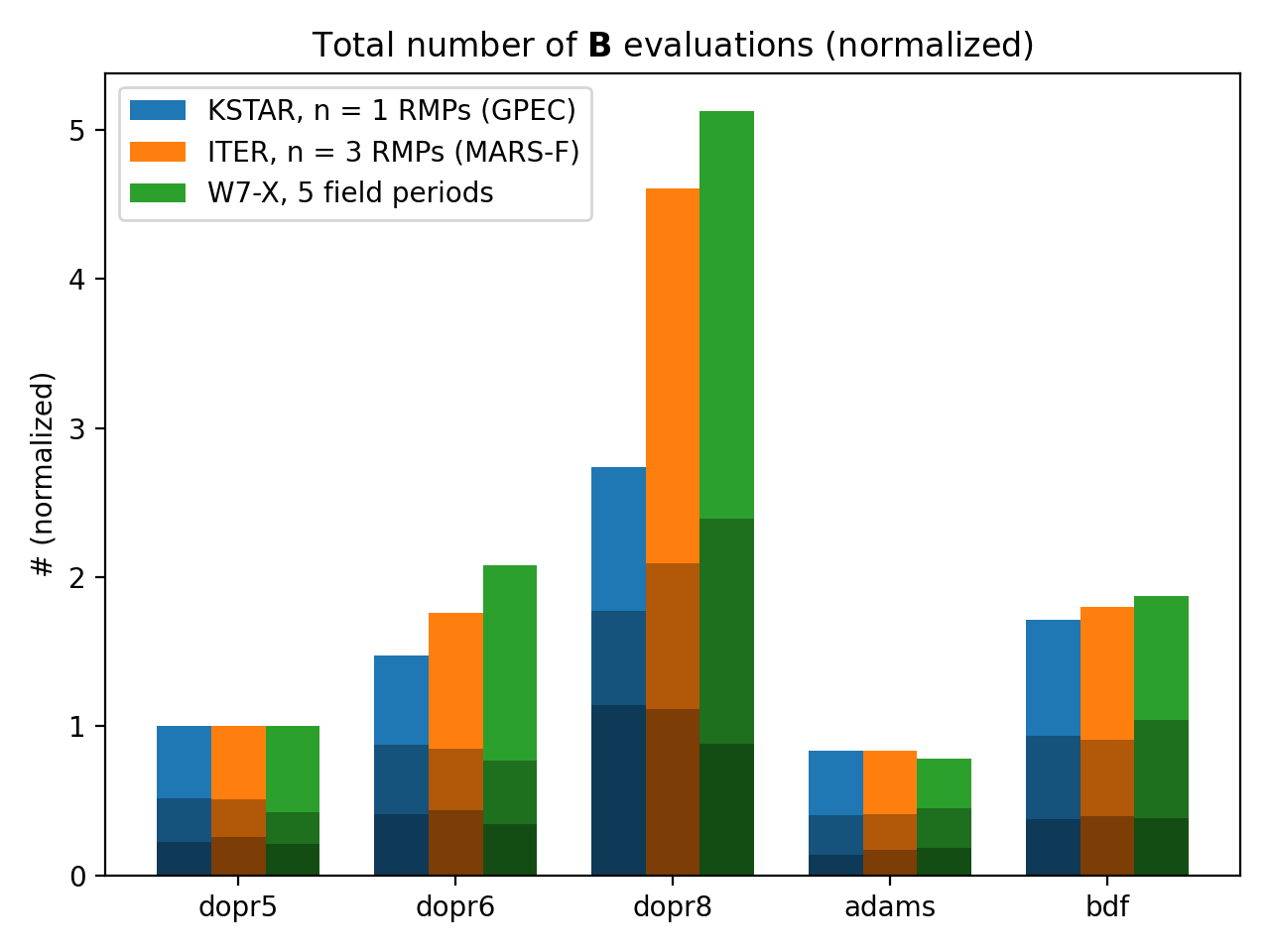}
\caption{Total number of $\vec{B}(\vec{r})$ evaluations in a magnetic footprint calculation with $\varepsilon = 10^{-7} \, \meter$ normalized to \texttt{dopr5}.
Three applications are shown: a KSTAR case \cite{Frerichs2023} ($n = 1$ RMPs with GPEC plasma response), an ITER case \cite{Frerichs2020, Frerichs2021a} ($n = 3$ RMPs with MARS-F plasma response), and a W7-X case \cite{Effenberg2019} (5 field periods).
Results from $\varepsilon = 10^{-6} \, \meter$ and $\varepsilon = 10^{-5} \, \meter$ are shown in comparison (consecutively darker colors).}
\label{fig:Bevaluations}
\end{figure}

Another aspect to consider is the efficiency of different integration methods.
Higher order methods typically come at the expense of more function evaluations $\vec{B}(\vec{r})$.
A measure for the efficiency of different integration methods for field line tracing is therefore the total number of $\vec{B}(\vec{r})$ evaluations (including ones from trail steps that are rejected during error control).
Figure \ref{fig:Bevaluations} shows that the \texttt{dopr5} method is the most efficient one of the Runge-Kutta methods for a given error tolerance of $\varepsilon = 10^{-7} \, \meter$.
In particular, the higher step accuracy of the \texttt{dopr6} and \texttt{dopr8} methods does not allow for a large enough increase of the step size in order to balance the number of additionally required function evaluations per step.
At a more relaxed error level of $\varepsilon = 10^{-5} \, \meter$, on the other hand, the \texttt{dopr8} method achieves similar global accuracy and efficiency as the \texttt{dopr5} method at $\varepsilon = 10^{-7} \, \meter$.
The \texttt{dopr6} method achieves a sufficiently good global accuracy of $\log_{10}\chi \, = \, -3.9$ at $\varepsilon \, = \, 10^{-6} \, \meter$ for the ITER case in figure \ref{fig:accuracy_benchmark} - and appears to perform slightly better than \texttt{dopr5} at $\varepsilon \, = \, 10^{-7} \, \meter$ and \texttt{dopr8} at $\varepsilon \, = \, 10^{-5} \, \meter$ for the case study in figure \ref{fig:Bevaluations}.

The variable order multi-step methods require only one $\vec{B}(\vec{r})$ evaluation for the predictor step and iteratively correct the initial guess.
Nevertheless, the \texttt{adams} method is only marginally more efficient than the \texttt{dopr5} method when error control is applied based on the same $\varepsilon$, and the \texttt{bdf} method turns out to be even less efficient.
The \texttt{adams} method achieves a good global accuracy of $\log_{10}\chi = -4.2$ already at $\varepsilon = 10^{-6} \, \meter$ and is therefore a good choice for the ITER case.
However, it only manages a moderate global accuracy of $\log_{10}\chi = -2.9$ at $\varepsilon = 10^{-7} \, \meter$ for the KSTAR case, which is less accurate than the \texttt{dopr5} method at the same $\varepsilon$.
Whether this can be attributed to the different perturbation model (and interpolation method) or to the different toroidal mode number remains unclear at this point.

\bibliographystyle{unsrt_doilink}
\footnotesize
\bibliography{references_copy}

\end{document}